\documentclass[sigconf]{acmart}
\AtBeginDocument{%
  }

\usepackage{booktabs}
\usepackage{array}
\usepackage{float}
\usepackage{enumitem}
\usepackage{subcaption}
\usepackage{graphicx} 
\usepackage{pdflscape}
\usepackage{tabularx}
\usepackage{booktabs}  
\begin{document}

%%
%% The "title" command has an optional parameter,
%% allowing the author to define a ''short title'' to be used in page headers.
\title[Anthropomorphism in Children's Interactions with LLM Chatbots]{Anthropomorphism in Children's Interactions with LLM Chatbots: A Systematic Review of Drivers and Outcomes}

\author{Hansinie Madushika Jayathilake}
\affiliation{%
  \institution{School of Information Technology}
  \institution{University of Cincinnati}
  \city{Cincinnati}
  \state{Ohio}
  \country{USA}
}
\email{jayathhm@mail.uc.edu}

\author{Renkai Ma}
\affiliation{%
  \institution{School of Information Technology}
  \institution{University of Cincinnati}
  \city{Cincinnati}
  \state{Ohio}
  \country{USA}
}
\email{mark@ucmail.uc.edu}

\renewcommand{\shortauthors}{Jayathilake et al.}

\begin{abstract}
Researchers across domains have investigated children’s use of LLM-based chatbots through various perspectives and methodologies. However, prior research remains fragmented regarding anthropomorphism, the tendency for children to assign human characteristics to those large language Model (LLM) chatbots as non-human objects. By analyzing 35 empirical studies published between 2022 and 2025, this systematic literature review identifies the drivers of anthropomorphism in children’s LLM chatbot interactions and the subsequent outcomes of these interactions. We found that human-like persona construction, adaptive scaffolding, supportive companionship, and non-human embodied design drive children’s anthropomorphic interactions. Additionally, five anthropomorphic outcomes emerged, including children exhibiting paradoxical social and moral responses, dual consciousness about the chatbots, forming varying social ties, exploring social boundaries, and attributing human narratives to conversation breakdowns. The findings, including both benefits and risks, can inform the future design and development of LLM chatbots focused on children’s well-being and promoting sustainable interactions that meet children’s developmental needs.
\end{abstract}




\maketitle

\section{INTRODUCTION}

Large language models (LLMs) have transformed how children interact with conversational agents, or chatbots, across domains of their daily lives. In educational settings, these agents support empathy development \cite{lo2025noel}, facilitate dialogic reading experiences \cite{he2025storypal}, scaffold creative coding activities \cite{druga2025scratch}, and enhance collaborative learning through timely peer interventions \cite{doherty2025piecing} for children. Beyond formal learning contexts, LLM chatbots serve as companions for storytelling \cite{chen2025characterizing, he2025storypal}, prompting children to share emotions about personal events \cite{seo2024chacha} and fostering curiosity-driven conversations that support well-being \cite{orancc2025talk}. In healthcare contexts, LLM chatbots create empathic spaces for teenagers to discuss mental health concerns \cite{marmol2025empathic}. Further, these chatbots foster environmental stewardship by enabling conversations about ecological crises \cite{tumedei2025drawings}, and facilitate family communication through integration with expressive arts therapy \cite{liu2024he}. This widespread adoption of LLM chatbots by children and families across education, health, and expression domains reflects LLMs' capacity to serve multifaceted roles in children's development and daily activities.

Such adoption of LLM chatbots also demonstrates anthropomorphism in children's interactions with them. Anthropomorphism refers to the attribution of human characteristics, motivations, intentions, and emotions to non-human entities, involving the perception of technological systems as possessing human-like qualities such as feelings, consciousness, or intentionality \cite{kim2023anthropomorphic, ma2025effect}. Children naturally anthropomorphize non-human entities as part of normal development, using them to make sense of their environment \cite{airenti2018development}. Piaget's theory suggests that children readily attribute intention and awareness to responsive, self-directed entities, which can make LLM-based chatbots be perceived as social and intentional agents \cite{piaget2017child}. LLM chatbots provide potent anthropomorphic cues through their conversational capabilities, such as peer-like design conversations \cite{lo2025noel}, responsive dialogue \cite{he2025storypal}, or empathetic communication \cite{marmol2025empathic}, so children might perceive these cues as genuinely social. The chatbots' capacity for natural language conversation, contextual memory, and self-disclosure might further reinforce such a tendency among children emotionally \cite{abercrombie2023mirages}.

Children's anthropomorphic interactions with LLM chatbots can be understood through the Epley et al.'s three-factor theory of anthropomorphism, which posits that anthropomorphism arises from three psychological determinants: \textit{elicited agent knowledge} (the accessibility of anthropocentric knowledge about the agent), \textit{effectance motivation} (the drive to understand and predict the agent's behavior), and \textit{sociality motivation} (the desire for social connection) \cite{epley2007seeing}. Prior work has applied this three-factor theory to understand anthropomorphism in human-robot interactions, examining appearance and agency \cite{crowell2019anthropomorphism}, social robot design incorporating dimensions of appearance, behavior, cognition, emotion, and morality \cite{david2022development}, and the role of anthropomorphism in motivating participation in citizen science projects \cite{rosen2024anthropomorphism}. Notably, prior research has applied the \citet{epley2007seeing}'s theory to examine children’s development through voice assistants \cite{festerling2022anthropomorphizing} and general child-agent interactions \cite{garg2022last}. However, this line of work remains largely conceptual and lacks a systematic synthesis of the specific drivers facilitating anthropomorphism within LLM-based chatbot interactions. To address this gap and identify the design elements that catalyze these attributions, we pose: \textbf{RQ1. What are the drivers in children's interactions with LLM chatbots that trigger or facilitate anthropomorphism?}

Recent systematic reviews have examined various outcomes of children's interactions with AI, including LLM chatbots. They have explored chatbot design effectiveness for learning and education \cite{ramandanis2023designing, alfarwan2025generative}, while \citet{feng2025effectiveness, park2025current}, and \citet{ozdemir2025digital} have examined chatbots' effectiveness in mental health management, More recently, research has begun investigating behavioral and social outcomes, including chatbots' role in bullying awareness and prevention \cite{lafrance2024uses}, and risk concerns in child-LLM interactions \cite{jiao2025llms}. However, a gap remains in synthesizing the anthropomorphic nature of these outcomes, specifically, how children's attribution of human traits to LLMs shapes their social and moral development. Systematically addressing this is essential for evaluating the long-term impact of these technologies, leading us to our second research question: \textbf{RQ2: What are the anthropomorphic outcomes of children's interactions with LLM chatbots?}

To answer these questions, we conducted a systematic literature review following PRISMA guidelines \cite{moher2009preferred}, analyzing 35 empirical studies published between 2022 and 2025. We identified four primary drivers of anthropomorphism: human-like persona construction that activates elicited agent knowledge; supportive companionship that satisfies sociality motivation; adaptive scaffolding that triggers effectance motivation; and non-human embodiment that extends these frameworks through fantasy archetypes (RQ1). We synthesized five anthropomorphic outcomes ranging from the formation of deep social ties and safe boundary exploration to the cognitive dissonance of dual consciousness and the ethical ambiguity of paradoxical social-moral responses (RQ2). 
To situate these findings, we move beyond a simple benefits-versus-risks binary to argue that children's anthropomorphic outcomes are deeply embedded within broader dimensions of their developmental wellbeing.

Our systematic literature review makes the following contributions to the Child-Computer Interaction (CCI) community: (1) a comprehensive systematic review of 35 empirical articles on child-LLM chatbot interactions, establishing a descriptive landscape of the field's current demographic, methodological, and contextual emphasis; (2) a taxonomy of anthropomorphic drivers developed using \citet{epley2007seeing}'s SEEK theory, extended to include non-human embodiment design as a fourth driver demonstrating that LLM conversational coherence alone is sufficient to trigger anthropomorphism independent of visual human-likeness \cite{festerling2022anthropomorphizing, gobel2025impact}; (3) a synthesis of five anthropomorphic outcome themes developed inductively from the included studies, moving beyond the benefits-versus-risks binary \cite{jiao2025llms, park2025current} to reveal how children's anthropomorphic engagement simultaneously implicates multiple dimensions of developmental wellbeing \cite{huppert2014state} in ways that vary qualitatively across Piagetian developmental stages \cite{piaget1964cognitive, hongneo}; and (4) a stakeholder-oriented future research and implications agenda identifying longitudinal investigation, stage-differentiated design, and the extension of child AI governance frameworks \cite{unicef2025ai, FiveRights2019} beyond safety thresholds as the three most pressing directions motivated by the findings.

\section{RELATED WORK}
\subsection{Theoretical Framework: The Three-Factor Theory of Anthropomorphism}
Anthropomorphism describes the psychological tendency to imbue the behavior of non-human agents with human-like characteristics \cite{epley2007seeing}. To explain this phenomenon, \citet{epley2007seeing} proposed the Three-Factor Theory (SEEK), identifying three psychological determinants driving anthropomorphism: \textit{\textbf{E}licited Agent Knowledge}, \textit{\textbf{E}ffectance Motivation}, and \textit{\textbf{S}ociality Motivation}. Elicited Agent Knowledge refers to the cognitive accessibility of anthropocentric knowledge; because individuals possess detailed self-knowledge, they use it to interpret unknown agents \cite{epley2007seeing}. Effectance Motivation describes the drive to attribute human traits to a non-human agent to reduce its uncertainty and helps users make sense and predict its behavior \cite{epley2007seeing}. Conversely, Sociality Motivation stems from the fundamental human need for connection; in the absence of human contact, individuals are more likely to anthropomorphize non-human agents to satisfy their desire for affiliation \cite{epley2007seeing}. 

We adopt the SEEK framework because it offers a robust structure for mapping specific LLM design features to the psychological drivers of interaction. Prior empirical research has primarily operationalized these factors to examine how specific design features trigger anthropomorphism in robotics. In social robotics, physical appearance and perceived agency act as primary stimuli for Elicited Agent Knowledge, where humanoid designs and autonomous behaviors trigger stronger anthropomorphic projections than non-humanoid or remotely controlled forms \cite{crowell2019anthropomorphism}. Recognizing that ``humanness'' is multidimensional, \citet{david2022development} developed the Social Robot Anthropomorphism Scale (SRA), identifying five distinct dimensions such as appearance, behavior, cognition, emotion, and morality that users evaluate during interaction. In the rapidly evolving context of Large Language Models (LLMs), \citet{xiao2025humanizing} argues that anthropomorphism is an intentional design concept mediated through perceptual, linguistic, behavioral, and cognitive cues that designers embed to guide user interpretation. Furthermore, \citet{gobel2025impact} highlights the role of Effectance Motivation in text-based interactions, finding that increased familiarity and frequency of use correlate with higher perceptions of human-likeness, as users build a predictive mental model of the chatbot over time.
 
Crucially, related to CCI, \citet{festerling2022anthropomorphizing}'s theoretical work argues that children’s anthropomorphism is not merely a ``playful error'' or an ``as-if'' pretense, but a developmental process shaped by their evolving social cognition. They suggest that as children interact with digital voice assistants exhibiting unprecedented combinations of human and non-human qualities, their experiential understanding of ``humanness'' shifts. However, while this framework establishes the developmental significance of such interactions, it predates the rise of generative AI (GenAI). Consequently, a systematic synthesis of recent empirical work on the drivers of anthropomorphism in the context of LLM chatbot interactions remains absent.

\subsection{Prior Reviews on Child–LLM Chatbot Interactions}
Recent studies examining children's interactions with LLM-powered chatbots have been largely focused on their functional efficacy in educational and therapeutic contexts, alongside critical safety assessments. In education, systematic reviews have evaluated chatbots as scalable pedagogical tools; \citet{ramandanis2023designing} analyzed development methods to highlight the utility of chatbots as conversational tutors, while \citet{alfarwan2025generative} noted the rapid integration of LLMs in K-12 settings to enhance learning engagement. Extending this into early childhood, \citet{ozdemir2025digital} found that AI-powered companions show promise in supporting social-emotional learning (SEL), specifically regarding self-awareness and emotional regulation though research remains fragmented. Parallel to education, a significant body of literature has examined the clinical role of these agents. \citet{park2025current} identified that many agents are designed as ``older peers'' to facilitate knowledge discovery on sensitive topics, yet often rely on rule-based prototypes with limited safety features. While \citet{feng2025effectiveness} reported that AI-driven agents demonstrated moderate effectiveness in reducing depressive symptoms, \citet{lafrance2024uses} highlighted critical limitations in crisis scenarios, noting that chatbots remain too predictable and inaccurate to effectively intervene in complex cases like youth bullying. Broader HCI reviews have also mapped how these agents support children’s play, storytelling, and identity formation \cite{garg2022last}, yet these opportunities are counterbalanced by severe ethical concerns regarding toxicity, bias, and data privacy \cite{jiao2025llms}. However, while these reviews map the functional landscape of benefits and risks, they often overlook how children’s cognitive maturity fundamentally constrains their understanding of these ``living'' or social entities, necessitating a closer examination of interaction through the lens of developmental psychology and a re-evaluation of outcome frameworks that move beyond safety thresholds to address children's wellbeing trajectories \cite{huppert2014state}.

\subsection{Developmental Context of Child–AI Interaction}
Children undergo distinct psychological developmental stages that shape their perception of the world, a process Jean Piaget described as the cognitive development of knowledge through active adaptation \cite{piaget1964cognitive}. According to Piaget, children progress through four major stages: the sensorimotor stage (birth to 2 years), where knowledge is acquired through physical interaction; the preoperational stage (2 to 6 years), characterized by symbolic thinking and egocentrism; the concrete operational stage (6 to 12 years), where logical thought emerges regarding tangible objects; and the formal operational stage (12 years and beyond), enabling abstract reasoning \cite{feldman2004piaget}. 

Recent work has begun to map these stages onto AI literacy. \citet{hongneo} proposed a Neo-Piagetian framework suggesting that users, particularly children, progress through parallel stages of AI understanding: preoperational users exhibit animistic thinking and treat the AI as a social peer; concrete operational users grasp the functional utility without understanding the mechanism; and formal operational users achieve abstract reasoning about the underlying computational architecture. \citet{hongneo} argues that unexpected AI behaviors, such as hallucinations or breakdowns, trigger cognitive conflict, which forces the user to accommodate their mental model, propelling them to a higher stage of understanding. This cyclical model implies that anthropomorphism is not a static error but a dynamic reflection of the user's current developmental stage \cite{hongneo}.

Despite the relevance of these developmental theories, the intersection of developmental psychology and empirical LLM research remains synthesis-poor. While prior reviews have cataloged the functional uses of chatbots in education and mental health \cite{alfarwan2025generative, park2025current}, and conceptual papers have proposed Neo-Piagetian frameworks for AI literacy \cite{hongneo}, no systematic review explicitly connects these developmental stages to the anthropomorphic drivers and outcomes in child-LLM engagements. Consequently, the HCI and CCI fields lack a cohesive understanding of how specific LLM chatbot design drives assimilation or accommodation processes across different Piagetian stages. Therefore, this systematic literature review aims to address this gap.

\section{METHODS}
To systematically investigate the drivers and outcomes of anthropomorphism in child-LLM chatbot interactions, we conducted a systematic literature review following the Preferred Reporting Items for Systematic Reviews and Meta-Analyses (PRISMA) guidelines \cite{moher2009preferred}. PRISMA is widely recognized for ensuring the transparency and reproducibility of review methodologies and has been extensively adopted in recent reviews on child-AI interaction (e.g., \cite{alfarwan2025generative,ramandanis2023designing, jiao2025llms, ozdemir2025digital, park2025current, lafrance2024uses}). Our review process adhered to the standard four-phase flow: (1) identification of records through database searching, (2) screening of titles and abstracts, (3) eligibility assessment via full-text review, and (4) inclusion of studies for data extraction using the inclusion criteria. 

\subsection{Identification: Inclusion Criteria}
Following established practices in prior systematic reviews in child-AI interactions \cite{ramandanis2023designing, park2025current} and HCI \cite{jiang2023trade, stowell2018designing}, we articulated the following inclusion criteria: 

\begin{itemize} 
    \item An article must be written in English to meet the language proficiency of the research team.
    
    \item An article must be published between 2022 and August 2025, the date we finished data collection. As we are interested in users' interactions with LLM chatbots, we selected this temporal criterion to align with the public launch of ChatGPT on November 30, 2022, which marked the advent of publicly and widely accessible LLM chatbots \cite{yu2023reflection}.
  
    \item An article must be a full-length empirical study. We define empirical as a study that collects and reports original primary data through a clearly stated methodology, such as controlled experiments, usability studies, field deployments, interviews, surveys, or observational studies. The study must involve direct or mediated interaction between child participants and a chatbot system, with findings derived from that data. System proposals, design explorations, conceptual frameworks, secondary analyses ,and non-full-text publication types such as abstracts, posters, editorials, and book chapters are excluded.
    
    \item An article must focus on a chatbot's target user group of children (individuals under 18), as we are interested in the anthropomorphic impact of LLM chatbots on young users. We exclude articles that focus on adult users or do not explicitly mention the chatbot's target user age.
 
  \item An article must explicitly identify the chatbot as being based on a Large Language Model (LLM) or Generative AI. We selected this technology criterion to distinguish our review from studies on older, rule based chatbots. We exclude articles that do not specify the underlying LLM model or focus on non-generative systems.

\end{itemize}

\subsection{Screening: Search Strategy}
To ensure a comprehensive coverage of relevant literature, we conducted a systematic search across three primary databases: ACM Digital Library, Scopus, and Web of Science. These databases were selected for their prominence in archiving high-impact research in HCI, learning sciences, and CCI \cite{alfarwan2025generative, garg2022last, lafrance2024uses}. Our search strategy combined keywords across three core categories, including target population, technology type, and interaction context. Keyword selection was informed by prior systematic reviews (e.g., \cite{garg2022last, ramandanis2023designing, alfarwan2025generative}), while technical nomenclature for conversational agents was derived from \citet{ramandanis2023designing}. The following Boolean search string was applied uniformly across all three databases:
("chatbot*" OR "conversational AI" OR "conversational agent*" OR "AI assistant*" OR "AI companion*") AND ("LLM" OR "large language model*" OR "generative AI" OR "artificial intelligence" OR "AI") AND (young OR youth OR "kid*" OR "teen*" OR "adolescent*" OR "child*"). The asterisk (*) denotes truncation to include all suffix variations of the root word (e.g., chatbot retrieves chatbots, chatbot's, etc.)
No database-specific query adaptations were made, as all three platforms supported standard Boolean operators and wildcard truncation. The complete search and screening procedure, detailing record attrition at each stage, is shown in Figure \ref{fig:prisma_diagram}.

\subsection{Eligibility Assessment}
\begin{figure*}[ht!]
    \centering
    \includegraphics[width=0.8\linewidth]{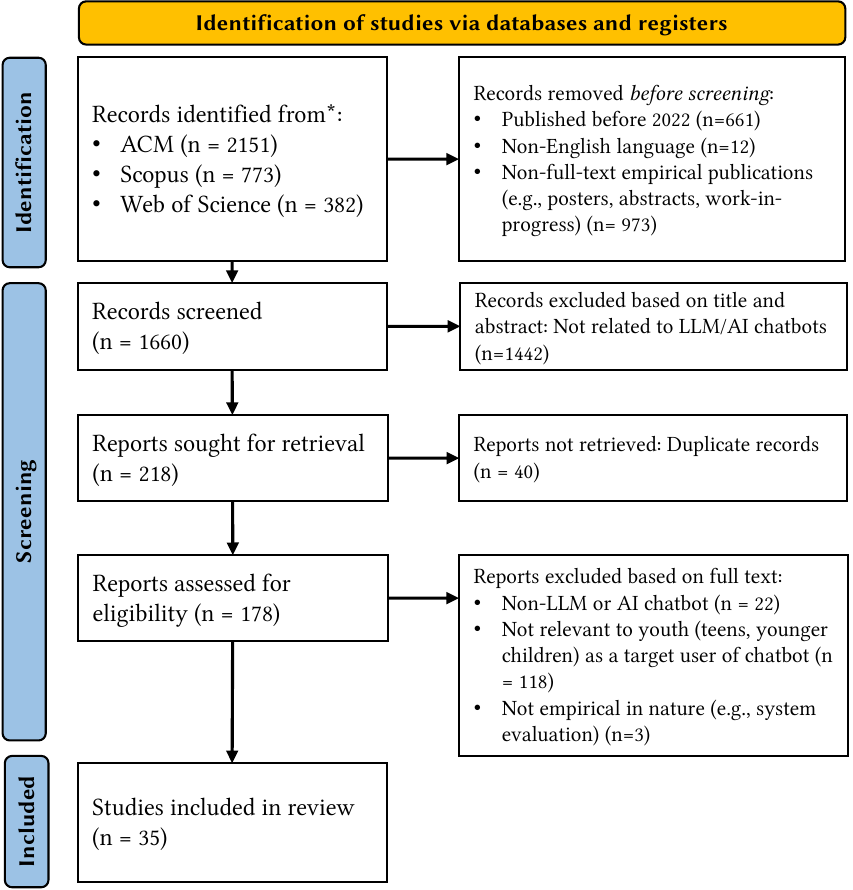}
    \caption{PRISMA flow diagram of the article selection process.}
    \label{fig:prisma_diagram}
    \Description{Diagrams showing the detailed process of study search selection aligned with PRISMA guidelines}
\end{figure*}

Following the initial screening of titles and abstracts, the first author assessed the full text of the remaining 178 articles for eligibility. To be included in the final review, studies were required to meet three eligibility criteria: (1) they must investigate a direct interaction between children (less than 18 years) and a chatbot system, (2) they must include a specific chatbot with a clear indication of the LLM model and (3) they must be empirical studies presenting original data with a clearly defined methodology. To ensure the reliability of the selection process, a second author verified the eligible studies against these criteria. Any discrepancies regarding inclusion were discussed between the two reviewers until a consensus was reached. Based on this dual assessment process, 143 studies were excluded. We categorized these exclusions as follows: (1) No LLM (N=22): Studies where the chatbot was not LLM-based (e.g., relied on rule based scripts or Wizard-of-Oz setups); (2) Population/Interaction Mismatch (N=118); (3) Articles that were not empirical in nature (n=3), such as conceptual frameworks, position papers, or system proposals lacking user evaluation. Consequently, 35 studies that met all eligibility criteria were selected for final data extraction and synthesis, as illustrated in Figure \ref{fig:prisma_diagram}.

\subsection{Inclusion: Data Extraction}

To ensure a consistent analysis of the 35 selected studies, we employed a standardized data extraction protocol. First, we extracted bibliographic and descriptive metadata to understand the field of child-LLM chatbot interactions, including publication year, title, venue, geographic location, target age group, specific use case (e.g., education, therapy), and the underlying LLM architecture (e.g., GPT-4, LLaMA). Following this, we adopted a template approach for qualitative analysis \cite{robson2024real} as used in prior CCI research \cite{lehnert2022child, van202018} to derive specific data points aligned with our research questions. For RQ1, we extracted excerpts related to the determinants of anthropomorphism, categorizing them into design cues (e.g., persona, voice) and interaction performance (e.g., personalized responses, error handling). For RQ2, we retrieved qualitative evidence of anthropomorphic outcomes, including verbatim quotes from children, behavioral observations, and reports from parents or experts regarding the child’s social engagement. To ensure reliability, the first author performed the initial extraction, which was then verified by the second author, with 100\% agreement, to ensure the extracted data accurately reflected the content of the included studies.

% \subsection{Methodological Profiling of Included Studies}
% To provide a transparent characterization of the evidence base, we conducted a methodological profiling of the 35 included studies following the approach recommended for qualitative evidence syntheses in systematic reviews \cite{popay2006guidance}. We profiled each study along six dimensions: (1) study design (e.g., quantitative, qualitative or mixed methods); (2) age range of child participants; (3) geographic context; (4) the specific LLM based chatbot evaluated; and (5) the primary outcome measured. No studies were excluded on methodological quality grounds at this stage, as our goal is a descriptive synthesis of an emerging evidence landscape rather than a cumulative effect estimate. The full methodological profile of all included studies is provided in Appendix Table \ref{tab:detailed-method-studies}

\subsection{Methodological Profiling of Included Studies}
To provide a transparent characterization of the evidence base, we conducted a methodological profiling of all 35 included studies. Following \citet{popay2006guidance}'s recommendation that data extraction in a systematic review should capture details of participants, interventions, outcomes, and study design to support assessments of applicability across contexts, we profiled each study along six dimensions: (1) research method (e.g., quantitative, qualitative, or mixed methods); (2) age range of child participants; (3) corresponding Piagetian developmental stage; (4) geographic context; (5) the specific LLM-based chatbot evaluated; and (6) publication venue. Consistent with \citet{popay2006guidance}'s guidance that reviewers pursuing descriptive synthesis may retain all studies while differentiating between them rather than excluding on methodological grounds, no studies were excluded at this stage, 
as our goal is a descriptive synthesis of an emerging evidence landscape rather than a cumulative effect estimate. The full methodological profile of all included studies is provided in Appendix Table \ref{tab:detailed-method-studies}.

\subsection{Data Analysis}

We employed a combined inductive-deductive qualitative approach \cite{newman2024want, strauss1987qualitative}, proceeding in three analytically distinct steps designed to ensure that themes emerged from the empirical evidence before theoretical frameworks were applied to organize them.

\subsubsection{Step 1: Preparatory Classification.} 
Prior to coding, two parallel classification activities were conducted to establish descriptive groundwork for the analysis.

Prior to coding, participant age ranges were extracted verbatim from each included study and mapped onto Piaget's developmental stages: Pre-operational (approximately 2–7 years), Concrete Operational (approximately 7–11 years), and Formal Operational (approximately 12 years and above) \cite{piaget1964cognitive}. Studies spanning multiple stages or lacking clearly defined age boundaries were assigned to all applicable categories. This classification was not used to generate codes but served as a developmental lens through which patterns emerging from Step 2 could subsequently be examined.

Each study was inductively assigned a primary use case label based on three boundary criteria applied in sequence: (1)the child's role in the interaction (consuming versus producing content); (2)whether the activity directed the child's attention toward their own feelings or toward external knowledge and others' perspectives; and (3)the deployment context (clinical or therapeutic versus educational or informal learning settings). Content domain alone was treated as insufficient for differentiation, as multiple studies share overlapping subject matter across functionally distinct purposes. Labels were iteratively refined until each study could be assigned to exactly one category without ambiguity. This classification was not used to generate thematic codes but provides a functional map of the evidence base that contextualizes the patterns identified in Steps 2 and 3.

\subsubsection{Step 2: Inductive Thematic Coding.} 
The first author developed initial codes directly from the extracted excerpts, drawing on the language used by researchers, children, parents, and practitioners in the included studies to describe observed interactions and responses. Codes were iteratively clustered into sub-themes and primary themes through constant comparison across studies. No pre-existing framework constrained code generation. For RQ1, coding focused specifically on which features of the chatbot's real-time interaction performance researchers reported as triggering anthropomorphic responses in children focusing dynamic behavioral qualities of the interaction rather than static design inventories such as the presence of a name or avatar. For RQ2, coding focused on the nature and character of children's responses, behaviors, and reported experiences during and after chatbot interaction. Outcomes were not coded as benefits or risks, which is most common in prior anthropomorphism literature \cite{akbulut2024all, placani2024anthropomorphism, abercrombie2023mirages}, but through what aspect of children's functioning and experience they implicated, allowing both positive and concerning expressions of the same phenomenon to be captured within a single theme. 

\subsubsection{Step 3: Deductive Theoretical Organization.}Following inductive coding, we examined whether the derived themes converged with established theoretical frameworks. For RQ1, we assessed whether inductively derived drivers mapped coherently onto \citet{epley2007seeing}'s three factors such as elicited agent knowledge, sociality motivation, and effectance motivation while retaining those that did and analyzing those that did not as inductively generated extensions. For RQ2, we examined convergence with Huppert's \cite{huppert2014state} ten-feature wellbeing model such as competence, emotional stability, engagement, meaning, optimism, positive emotion, positive relationships, resilience, self-esteem, and vitality by checking whether the behavioral descriptions and verbatim quotes in the included studies used language congruent with specific flourishing features. 

To ensure analytical rigor, the first and second authors engaged in multiple iterations of collaborative discussion to negotiate, revise, and finalize the themes until consensus was reached \cite{hill2005consensual}. This approach, involving a single primary coder who regularly meets with the research team for discussion and refinement, is a common and rigorous practice in prior HCI work \cite{jiang2021supporting, patel2019feel}.

To ensure traceability, each sub-theme in the results tables is accompanied by citations to contributing studies, allowing readers to verify every claim against appendix  Table \ref{tab:detailed-method-studies}.

\section{FINDINGS}
This section presents how previous literature was analysed based on the children's anthropomorphic interactions with LLM chatbots. It will provide a description of the prior work (Section 4.1), the drivers of anthropomorphism in chatbot design (Section 4.2), and the anthropomorphic outcomes of children with LLM chatbots (Section 4.3).

\subsection{Descriptive Results of Prior Studies on Child-LLM Chatbot Interactions}
\subsubsection{Publication Trends and Venues.} Research interest in child-LLM chatbot interaction has surged exponentially, growing from a single paper in 2023 to 15 in 2024, and reaching 19 in the first quarter of 2025 alone. The discourse is highly centralized in HCI venues, primarily the ACM Conference on Human Factors in Computing Systems (CHI, N=17) and the ACM Interaction Design and Children Conference (IDC, N=7). The remaining contributions are distributed across the ACM Designing Interactive Systems Conference (DIS, N=2) and single publications (N=1) in the International Conference on Information and Communication Technologies and Development (ICTD), National Council on Family Relations (NCFR), International Conference on Tangible, Embedded, and Embodied Interaction (TEI), International Journal of Human Computer Interaction (IJHCI), Frontiers in Artificial Intelligence, Reading Research Quarterly (RRQ), International Conference on E-Education, E-Business, E-Management and E-Learning (IC4E), Institute of Electrical and Electronics Engineers (IEEE), and Association for Computing Machinery (ACM).

\begin{figure*}[ht!] 
    \centering
    % Row 1
    \begin{subfigure}[b]{0.48\textwidth}
        \centering
        \includegraphics[width=\textwidth]{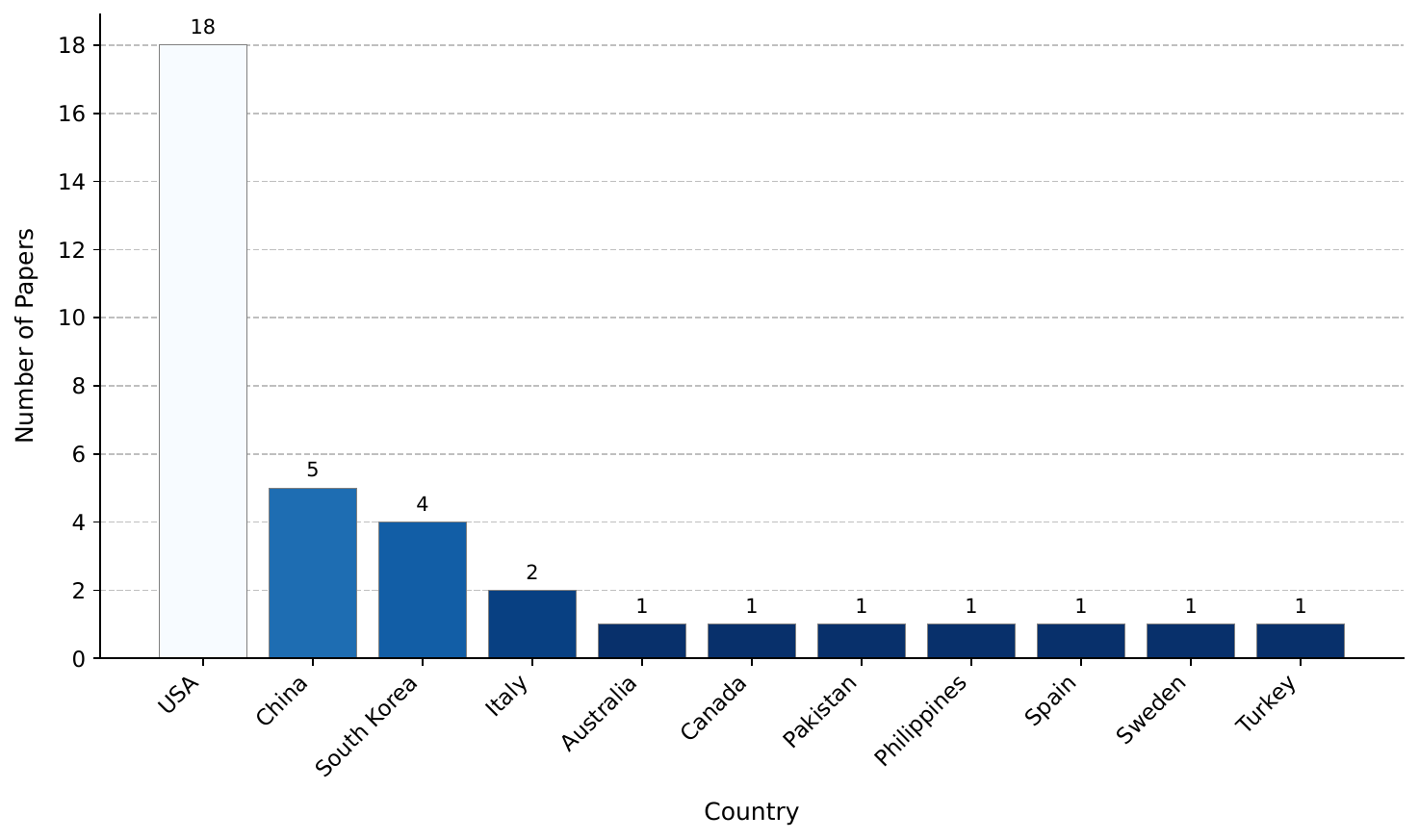}
        \caption{Geographic Distribution}
        \label{fig:country}
    \end{subfigure}
    \hfill
    \begin{subfigure}[b]{0.48\textwidth}
        \centering
        \includegraphics[width=\textwidth]{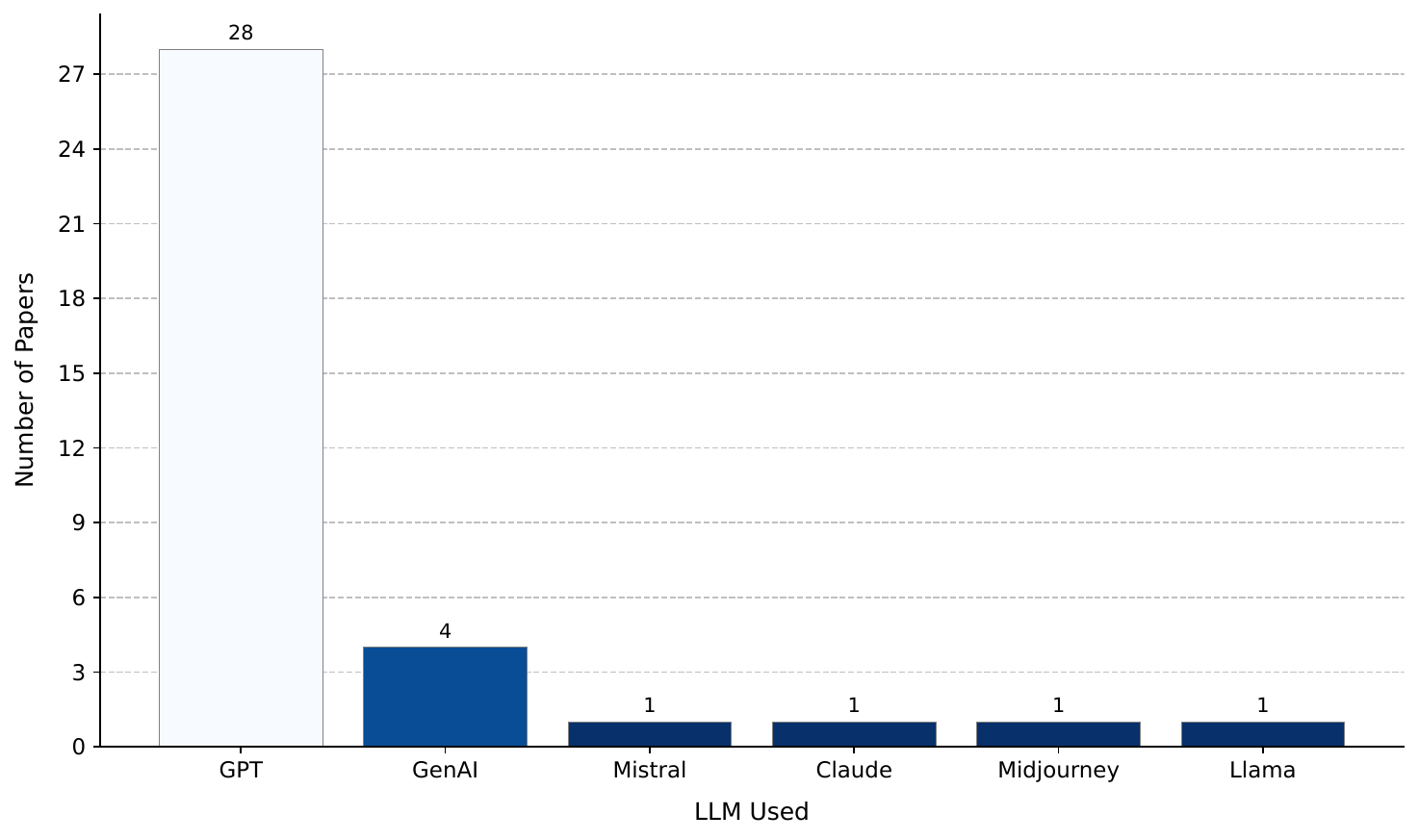}
        \caption{LLM Model Adoption}
        \label{fig:llm}
    \end{subfigure}
    
    \vspace{0.5cm} % Spacing between rows
    
    % Row 2
    \begin{subfigure}[b]{0.48\textwidth}
        \centering
        \includegraphics[width=\textwidth]{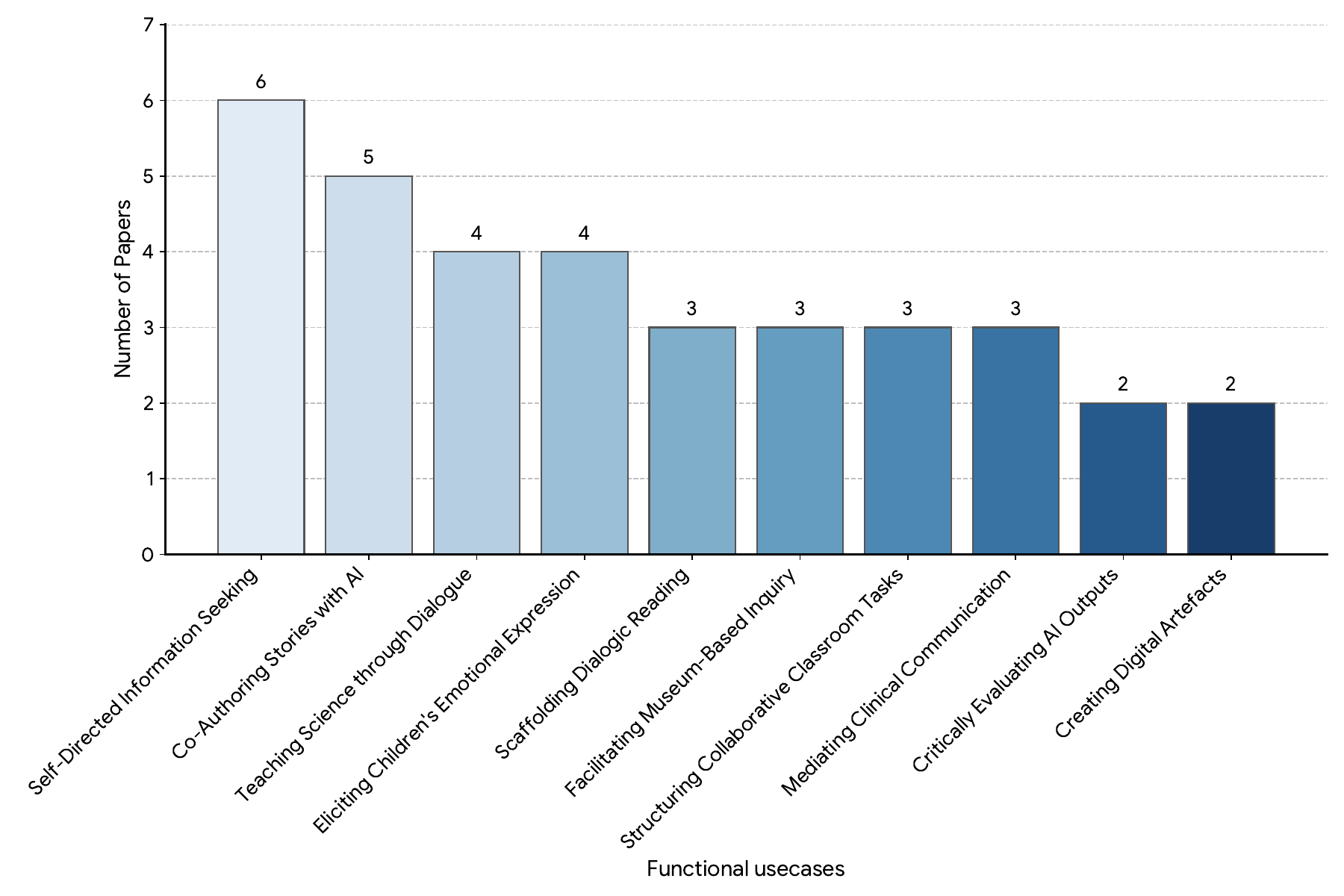}
        \caption{Functional Use Cases}
        \label{fig:usecase}
    \end{subfigure}
    \hfill
    \begin{subfigure}[b]{0.48\textwidth}
        \centering
        \includegraphics[width=\textwidth]{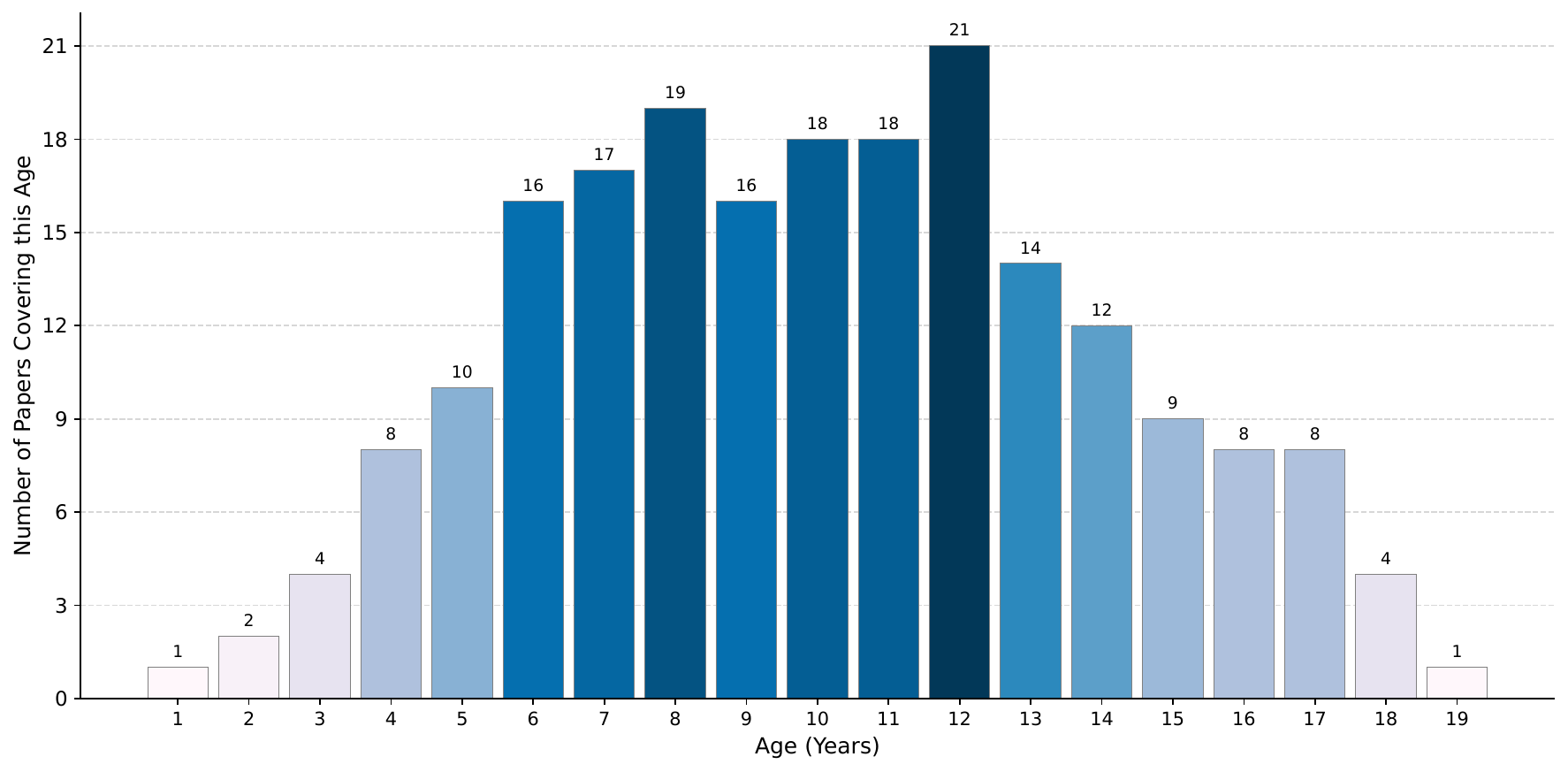}
        \caption{Target User Age Distribution}
        \label{fig:age}
    \end{subfigure}
    
    \caption{Descriptive statistics of the reviewed articles (N=35): (a) Distribution by country of origin (determined by the first author's affiliation address), (b) Underlying LLM model choice of LLM chatbots, (c) Primary functional use case, and (d) Target user age (to reflect the full developmental span, studies targeting an age range were coded repetitively for every single year included; e.g., a study focusing on ages 5–8 was coded four times, once each for ages 5, 6, 7, and 8).}
    \label{fig:descriptive_stats}
    \Description{A set of diagrams showing the current statistics for included studies in the review.}
\end{figure*}

\subsubsection{Geographic, Technological, Demographic and Methodological Contexts} Geographically, the field is predominantly led by the USA (N=18), with emerging interest from East Asia, specifically China (N=5) and South Korea (N=4) (Figure \ref{fig:descriptive_stats}a). Technologically, a distinct ``monoculture'' exists; 28 studies utilized OpenAI's GPT models, vastly overshadowing alternatives like Mistral or Claude (N=1 each) (Figure \ref{fig:descriptive_stats}b). Functional applications are led by Self-Directed Information Seeking (N=6), followed by Co-Authoring Stories with AI (N=5), Teaching Science through Dialogue (N=4) and Eliciting Children's Emotional Expression (N=4) (Figure \ref{fig:descriptive_stats}c). 
Regarding target demographics of LLM chatbots, the age distribution of children participants, coded for each specific year reported in each study, follows a bell curve, with a salient concentration in the Concrete Operational stage (ages 6–12 years). The transition into the Formal Operational stage (age 12) represents the single most studied cohort (N=21). In contrast, the Sensorimotor and Preoperational stages (birth to 6 years) and late adolescence (ages 17+) remain under-explored (Figure \ref{fig:descriptive_stats}d).

Methodologically, the majority of included studies employed mixed methods designs (n=19), followed by qualitative approaches (n=13) and quantitative approaches (n=3), reflecting the field's predominantly exploratory yet empirically grounded orientation toward child-LLM based chatbot interaction. Regarding study duration, most studies were short-term, involving single-session or brief multi-session deployments, with only a small number involving longitudinal or repeated-contact designs \cite{shi2025need}. The design goals of the evaluated chatbots were predominantly educational, spanning use cases from literacy scaffolding and science dialogue to creative co-authoring and collaborative problem solving (Figure \ref{fig:descriptive_stats}c), with a smaller cluster targeting clinical or therapeutic communication contexts. This short-term, educationally oriented evidence base has direct implications for interpreting the RQ2 findings in which the anthropomorphic outcomes identified in this review primarily reflect children's immediate interactional responses rather than sustained developmental trajectories, a limitation discussed further in Section 5.4.

\subsection{RQ1: Drivers of Anthropomorphism in Child-LLM Chatbot Interactions}
Our analysis identified four primary drivers that facilitate children's anthropomorphic responses to LLM chatbots, three of which align with \citet{epley2007seeing}'s three-factor theory and one that extends beyond it (See Table \ref{tab:rq1_Drivers}). Specifically, we found that LLM chatbots drive anthropomorphism by activating children's elicited agent knowledge through \textit{Human-Like Persona Construction}, satisfying sociality motivation via \textit{Supportive Companionship}, engaging effectance motivation through \textit{Adaptive Scaffolding}, and leveraging \textit{Non-human Embodiment Design} to make fantasy characters appear alive by synchronizing their speech, movements, and visuals, as summarized in Table \ref{tab:rq1_Drivers}.

\subsubsection{Human-Like Persona Construction} This driver, (n=19), activates children's elicited agent knowledge by presenting human-like characteristics and behaviors through five primary persona designs. (1) Human-like roleplay, created detailed characters or personas with backstories such as Noel, a 12 year old with low vision \cite{lo2025noel}, and peers with cancer \cite{seo2025enhancing}, along with vulnerable personas that acted as adolescents with worries to encourage reciprocal self-disclosure \cite{marmol2025empathic}, while \citet{marmol2025empathic,park2025beyond} offered multiple persona selection options based on gender or distinct agent roles. (2) Voice and Auditory Human-Like Cues, provided gendered and accented voice characteristics \cite{tang2024emoeden, vella2025hello, tumedei2025you} and voice-based interaction modalities \cite{he2025storypal,tumedei2025drawings,oh2025hey,shi2025need}. (3) Visual and physical expression cues displayed emotional expression through personalized emojis and facial expressions \cite{tang2024emoeden} and physical behavioral displays representing cognitive states \cite{oh2025hey, orancc2025talk}. (4) Communicative competence featured first-person singular pronouns \cite{oh2025hey,tumedei2025you}, sustained dialogue capability \cite{marmol2025empathic, doherty2025piecing, belghith2024testing,park2025beyond}, and accurate input processing with 87\% utterance recognition \cite{belghith2024testing,zhang2024mathemyths}. Demonstrated cognitive capabilities, (mental processes) with human-comparable creative performance \cite{zhang2024mathemyths}, contextual intelligence and inference \cite{belghith2024testing, zhang2025qualitative, tumedei2025you}, and progressive human-likeness \cite{torrato2024knowledge, yu2025exploring}, making more teenagers turn to chatbots. These human-agent cues activated children's readily accessible anthropocentric knowledge, making it highly applicable to the AI agent and facilitating anthropomorphic attribution.

\subsubsection{Supportive Companionship} We found that this driver (n=16) facilitates anthropomorphism by satisfying children's sociality motivation, the innate drive to connect with others, through the construction of safe, emotionally responsive, and playful identities. Specifically, this driver operates through three key mechanisms. First, presentation of Supportive Companionship (n=16) involved friendly companion self-presentation as ``your AI friend''\cite{druga2025scratch, belghith2024testing} with playful first-person representation \cite{tumedei2025you}, and safe peer level relational positioning where children felt more comfortable since chatbots were ``unbiased, nonjudgmental, accessible'' \cite{seo2025enhancing} and parents claim ``at least safer than those harmful people'' \cite{han2024teachers}. Second, supportive Emotional Responsiveness demonstrated emotional understanding through reflective empathetic responses \cite{seo2024chacha} and humanized expressions to evoke empathy \cite{tang2024emoeden}, while providing encouragement through positive feedback libraries \cite{shi2025need}, interactive encouragements with emojis \cite{zha2025mentigo}, and child friendly supportive language \cite{dangol2025ai}. Last, Entertaining Role Adoption featured playful communication with humor \cite{vella2025hello, tumedei2025you}, flexible interaction modes offering task-oriented and social oriented approaches \cite{lo2025noel} and third player roles \cite{zhang2025qualitative}, and casual social conversation through free form creative conversations \cite{yu2025exploring}, open child-parent communication \cite{li2024said}, and imaginative playful scenarios \cite{torrato2024knowledge}. These companionship features satisfied children's need for social connection by enabling perceived human-like relationships, particularly appealing when children sought non-judgmental partners, activating sociality motivation.

\subsubsection{Adaptive Scaffolding}  
We found that this driver, (n=14) facilitates anthropomorphism by activating children's effectance motivation, the drive to master their environment, by reducing interactional uncertainty and enhancing competence through responsive mentoring. This operates through two distinct mechanisms. First, Cognitive Scaffolding for Skill Acquisition (n=12) employed strategic question generation that dynamically adjusted support and strategically increased question complexity \cite{he2025storypal, chen2025characterizing,torrato2024knowledge, dietz2024contextq, ly2025museum}, along with comprehension support techniques including adaptive explanations \cite{lee2023dapie}, voice-over playback controls \cite{tang2024emoeden}, word limit constraints \cite{tumedei2025you}, progressive personalization reducing overlap from 51\% to 10\% \cite{choi2025aacesstalk}, and customization of content and interaction modes \cite{klarin2024adolescents, choi2025aacesstalk}. Second, Affective Scaffolding provided scaffolding through emotional support maintaining encouraging tones with graduated assistance \cite{he2025storypal}, open-ended guidance questions when hesitant \cite{shi2025need}, and explicit encouragement for silent children \cite{he2025storypal}, along with adaptive responsiveness that encouraged step-by-step task completion \cite{zha2025mentigo}, responded to low motivation \cite{marmol2025empathic}, dynamically identified over 23 distinct student states \cite{zha2025mentigo}, and asked questions to develop interest \cite{marmol2025empathic,druga2025scratch}. By providing responsive assistance that reduced uncertainty and increased confidence, these scaffolding features enhanced children's ability to interact effectively, activating effectance motivation and facilitating anthropomorphic attribution of intentional teaching characteristics.

\begin{table*}[t]
\scriptsize
\centering
\renewcommand{\arraystretch}{1.4}
\caption{Anthropomorphic drivers in child-LLM interactions (RQ1). The values in parentheses (n=xx) represent the number of studies identifying each theme out of the total (N=35).}
\label{tab:rq1_Drivers}
% We use tabularx to fill the full page width
\begin{tabularx}{\textwidth}{@{} p{2.2cm} p{4cm} p{2.8cm} X @{}}
\toprule
\textbf{Primary Theme} & \textbf{Theme explanations} & \textbf{Sub-Theme} & \textbf{Specific drivers} \\ 
\midrule

% ROW 1: Human-Like Persona
Human-Like Persona Construction \newline \textit{n=19} & 
Activates \textit{elicited agent knowledge} by providing cues that map directly onto children's pre-existing schemas of human identity, communication, and cognition. & 
Human-like Roleplay & 
Detailed character personas with backstories \cite{lo2025noel, tumedei2025you, marmol2025empathic, seo2025enhancing}; Multiple persona selection options \cite{marmol2025empathic, park2025beyond}; Peer-like behavioral positioning \cite{seo2024chacha, marmol2025empathic, orancc2025talk, seo2025enhancing} \\
\cmidrule{3-4} 
 & & Voice and Auditory Cues & 
Gendered and accented voice characteristics \cite{tang2024emoeden, vella2025hello, tumedei2025you}; Voice-based interaction modality \cite{he2025storypal, tumedei2025drawings, oh2025hey, shi2025need} \\
\cmidrule{3-4}
 & & Visual Expression Cues & 
Emotional expression through visuals \cite{tang2024emoeden}; Physical state and behavioral displays \cite{oh2025hey, orancc2025talk} \\
\cmidrule{3-4}
 & & Communicative Competence & 
First-person singular pronoun (``I'') \cite{tumedei2025you, oh2025hey}; Sustained dialogue capability \cite{doherty2025piecing, marmol2025empathic, belghith2024testing, park2025beyond}; Accurate input processing \cite{belghith2024testing, zhang2024mathemyths} \\
\cmidrule{3-4}
 & & Cognitive Performance & 
Creative Performance \cite{zhang2024mathemyths}; Contextual Inference \cite{zhang2025qualitative, tumedei2025you, belghith2024testing}; Progressive human-likeness \cite{torrato2024knowledge, yu2025exploring}\\
\midrule

% ROW 2: Supportive Companionship
Supportive Companionship \newline \textit{n=16 } & 
Satisfies \textit{sociality motivation} by establishing a safe, non-judgmental partnership that fulfills children's psychological need for connection and emotional validation. & 
Safer Companionship & 
Friendly companion self-presentation \cite{tumedei2025you, druga2025scratch, belghith2024testing}; Safe peer-level relational positioning \cite{seo2024chacha, han2024teachers, seo2025enhancing} \\
\cmidrule{3-4}
 & & Emotional Responsiveness & 
Demonstrating emotional understanding \cite{tang2024emoeden, seo2024chacha}; Encouragement through positive tone \cite{dangol2025ai, zha2025mentigo, shi2025need} \\
\cmidrule{3-4}
 & & Entertaining Role Adoption & 
Playful entertaining communication \cite{vella2025hello, tumedei2025you}; Flexible social interaction modes \cite{lo2025noel, zhang2025qualitative}; Casual social conversation \cite{torrato2024knowledge, yu2025exploring, li2024said} \\
\midrule

% THEME 3
Adaptive Scaffolding \newline \textit{n=14 } & 
Triggers \textit{effectance motivation} by reducing interactional uncertainty and enhancing children's competence through responsive, mentor-like guidance. & 
Cognitive Scaffolding &
Strategic question generation \cite{he2025storypal, chen2025characterizing, torrato2024knowledge, dietz2024contextq, ly2025museum}; Comprehension support techniques \cite{tang2024emoeden, tumedei2025you, klarin2024adolescents, choi2025aacesstalk, chen2025characterizing, lee2023dapie} \\
\cmidrule{3-4}
 & & Affective Scaffolding & 
Emotional support and patience \cite{he2025storypal, shi2025need}; Adaptive responsiveness to sustain engagement \cite{marmol2025empathic, druga2025scratch, zha2025mentigo} \\
\midrule

% THEME 4
Non-human Embodiment Design \newline \textit{n=10 }  & 
Extends \textit{elicited agent knowledge} by leveraging fantasy archetypes to bridge the gap between non-human appearance and coherent social behavior. & 
Non-human Character &
Whimsical Animal Characterization \cite{chen2025characterizing, orancc2025talk}; Nature Embodiment \cite{vella2025hello, tumedei2025drawings}; Cartoon-style Visual Design \cite{tang2024emoeden, vella2025hello, oh2025hey, han2024aistory} \\
\cmidrule{3-4}
 & & Synchronized Sensory Coordination & 
Coordinated Speech-Text Presentation \cite{lee2023dapie, park2025beyond, han2024aistory}; Physical-Digital Integration \cite{tang2024emoeden, shi2025need, park2025beyond, orancc2025talk} \\
\bottomrule
\end{tabularx}
\end{table*}

\subsubsection{Non-human Embodiment Design} This driver,( n=10), extends beyond Epley's SEEK framework by paradoxically facilitating anthropomorphism through fantasy characters with integrated technological coordination. This effect is achieved through two main strategies. First, Non-human Character Embodiment featured whimsical animal characterization such as emus chosen for their ``slightly humorous and distinctive appearance'' \cite{orancc2025talk} and penguin characters \cite{chen2025characterizing}, nature embodiment as lagoons or friendly trees \cite{vella2025hello,tumedei2025drawings}, and cartoon style visual design with rounded shapes \cite{tang2024emoeden}, cartoon faces \cite{vella2025hello}, and gender-neutral cartoon cats \cite{oh2025hey,han2024aistory}. Synchronized Sensory Coordination (n=4) provided coordinated speech-text presentation \cite{lee2023dapie, park2025beyond,han2024aistory} and physical-digital integration through sock puppets mimicking animations \cite{vella2025hello}, touch-based interactions \cite{orancc2025talk}, and camera-based activity recording \cite{shi2025need,park2025beyond}. Rather than reducing anthropomorphism, this driver activated the elicited agent knowledge about fantasy characters while creating cohesive sensory experiences. Such multimodal coordination signaled intentional communication and responsive awareness, facilitating anthropomorphic attribution despite explicitly non-human visual presentation, suggesting that coherent cross-modal responsiveness may override surface appearance in activating anthropomorphic responses.

\subsection{RQ2: Anthropomorphic Outcomes of Child-LLM Chatbot Interactions}
We identified five primary outcomes that emerge from children’s anthropomorphic engagement with LLM chatbots. These outcomes reflect a complex interplay where children construct deep social bonds, explore emotional boundaries, and navigate the cognitive dissonance of interacting with a ``hybrid'' entity that they consider both friend and machine, as summarized in Table \ref{tab:outcome_taxonomy}.

\subsubsection{Social Ties Construction} The most pervasive outcome (n=20) is Social Ties Construction, where children actively build distinct social relationships with chatbots, ranging from deep emotional attachments to functional partnerships. \citet{granovetter1983strength}, distinguished social relationships as strong ties close relationships with frequent, emotionally dense interactions (ties with family and close friends) and weak ties which involve relationships with less intense interactions like casual contacts. In this context, children form Strong Emotional Ties by attributing profound roles to the agent; they perceive chatbots as ``Parental Figures'' capable of sharing burdens \cite{shi2025need, lee2023dapie}, ``Romantic Partners'' offering stability and fantasy fulfillment \cite{yu2025exploring}, or ``Close Friends'' for sharing secrets \cite{seo2024chacha,seo2025enhancing}. Gendered attributions, referring to the bot as ``he'' or ``she'' based on voice or helpfulness—further solidify these bonds \cite{he2025storypal,belghith2024testing,orancc2025talk}. On the other hand, these ties can be more functional, manifesting as Weak Instructional Ties where the bot is viewed as a ``Teacher'' or ``Mentor'' valued for intelligence and effectiveness \cite{lee2023dapie}, or Weak Entertainment Ties where the agent serves as a ``Playful Co-Participant'' or ``Teammate'' in games \cite{zhang2025qualitative,dangol2025ai,druga2025scratch}. This spectrum illustrates that anthropomorphism is not a monolith; children tailor the social tie intimate or instrumental based on the perceived affordances of the specific agent.

Social tie formation with LLM chatbots spans all three Piagetian stages, yet the character of those ties shifts in developmental significance across sub-themes. Strong emotional ties including parental figure attribution \cite{he2025storypal, shi2025need, lee2023dapie}, peer and friendship bonds
\cite{seo2025enhancing, belghith2024testing, higgs2024being, zha2025mentigo, marmol2025empathic, seo2024chacha, druga2025scratch, orancc2025talk}, and gendered companion roles \cite{he2025storypal, belghith2024testing, orancc2025talk, vella2025hello} are documented across preoperational, concrete operational, and formal operational stages, indicating that relational attribution to chatbots emerges in early childhood and persists through adolescence. however, romantic partner relationships \cite{yu2025exploring}, are the most developmentally specific outcome in this
sub-theme, appearing exclusively at the formal operational stage, which suggests that the most concerning form of substitute attachment which frames a chatbot as a romantic partner requires the abstract social cognition characteristic of adolescence. Weak instructional ties \cite{yu2025exploring, seo2025enhancing, lee2023dapie, chen2025characterizing, zha2025mentigo, druga2025scratch, ly2025museum, belghith2024testing} and entertainment ties \cite{druga2025scratch, zhang2025qualitative, orancc2025talk, park2025beyond, dangol2025ai,
tumedei2025drawings, khan2024chatgpt, higgs2024being, vella2025hello, chen2025characterizing, he2025storypal} are similarly present across all three stages, indicating these more functional relational roles are developmentally accessible from preschool onward and represent the broadest and most stage-inclusive expressions of social tie formation in this theme.

\subsubsection{Social Boundary Exploration} A second critical outcome (n=10) is Social Boundary Exploration, where children utilize the LLM chatbot as a safe, non-judgmental space to test social and emotional limits. On one hand, this manifests as Emotional Security, where children engage in intimate disclosure, sharing secrets, family conflicts, or even suicidal thoughts with the bot because it listens without the risks associated with human judgment \cite{seo2024chacha,marmol2025empathic,yu2025exploring}. This sense of safety often leads to flourishing conversations where children feel confident discussing mental health or personal interests \cite{marmol2025empathic}. On the other hand, children actively seek Social Validation through interrogative behaviors. They probe the chatbot with acquaintance questions about its age, favorites, or happiness \cite{park2025beyond,tumedei2025drawings} and sometimes test its loyalty by checking if it remembers past details \cite{vella2025hello}. While this safe space fosters openness, it can also lead to frustration when the ``friend'' becomes repetitive or fails to recall shared history, breaking the illusion of a unique social bond \cite{seo2024chacha}.

Social boundary exploration shows the clearest developmental distinction of any outcome theme in the table. Social validation behaviors including acquaintance questions
\cite{park2025beyond, belghith2024testing, vella2025hello, tumedei2025drawings}, preference questions \cite{seo2024chacha, park2025beyond, tumedei2025drawings}, empathetic inquiries \cite{seo2024chacha, tumedei2025drawings}, and knowledge verification \cite{oh2025hey, belghith2024testing} span all three stages, indicating that interrogative social-probing of chatbots begins in early childhood and persists through adolescence. Emotional security behaviors, however, are absent from the preoperational stage entirely: intimate disclosure \cite{seo2024chacha, marmol2025empathic, yu2025exploring}, flourishing conversations
\cite{lo2025noel, marmol2025empathic}, and frustration with routine \cite{seo2024chacha, seo2025enhancing} all concentrate in concrete operational and formal operational stages only. This pattern indicates that the emotionally consequential dimension of social boundary exploration including disclosing personal information, experiencing genuinely enriching conversations, and feeling frustrated when relational expectations are unmet requires at least concrete operational social cognition. The developmental asymmetry between social validation, which is broadly accessible, and emotional security, which is developmentally bounded, suggests that younger children engage in the surface form of boundary-probing without the depth of emotional investment that makes the same behavior developmentally significant for older children.

\subsubsection{Human Narrative Attribution} When interactions break down, children employ Human Narrative Attribution (n=12) as a repair mechanism, projecting human states onto the machine to make sense of failure. On one hand, children explain technical glitches through Physical and Psychological States, attributing silence to the bot being ``hungry'', ``tired'', ``scared'',or even ``dead'' \cite{vella2025hello,seo2024chacha}. They may frame the bot’s inability to answer as a human-like deficiency, such as being stupid or ``not listening'' \cite{dangol2025ai}. On the other hand, this attribution enables Repair Behaviors where children use conversation repair techniques like simplifying language or using visual metaphors, assuming the ``friend'' just needs a better explanation \cite{dangol2025ai,li2024said}. While this allows children to celebrate success when the bot finally understands \cite{dangol2025ai}, persistent failures that shatter the human narrative can lead to relationship abandonment, where the child rejects the agent entirely out of frustration \cite{dangol2025ai,newman2024want}.

Human narrative attribution as a cognitive repair mechanism is documented across all three Piagetian stages in every sub-theme, yet the distribution within sub-themes reveals meaningful developmental differentiation. Physical and psychological state attributions including human physical characteristics \cite{vella2025hello} and cognitive and emotional deficiencies \cite{he2025storypal, seo2024chacha, dangol2025ai, belghith2024testing} span preoperational through formal operational stages, suggesting that projecting human internal states onto chatbot failures is a cross-stage phenomenon, though its expression likely differs from animistic physical attribution in younger children to more psychologically nuanced deficiency framing in older ones. Dispositional critiques including character attribution \cite{higgs2024being, druga2025scratch, ly2025museum, newman2024want, dangol2025ai, zhang2025qualitative, belghith2024testing, tumedei2025you}, humor in failures \cite{dangol2025ai}, and acceptance and compromise \cite{dangol2025ai, liu2024he, tang2024emoeden, lo2025noel, vella2025hello, newman2024want} are similarly distributed across all three stages, indicating that evaluative responses to chatbot failure, from finding it amusing to accommodating imperfection, are not developmentally bounded. Repair behaviors reveal the most notable within-theme pattern: while conversation repair techniques \cite{li2024said, dangol2025ai, oh2025hey} and celebration of success \cite{dangol2025ai, belghith2024testing} appear across preoperational through formal operational stages, relationship abandonment \cite{dangol2025ai, newman2024want} is confined to the concrete operational stage, suggesting that the decisive behavioral rupture with a chatbot following accumulated failure is most characteristic of middle childhood rather than either earlier or later developmental periods.

\begin{table*}[t]
\scriptsize
\centering
\renewcommand{\arraystretch}{1.4}
\caption{Outcomes of child-LLM chatbot interactions (RQ2). The values in parentheses (n=xx) represent the number of studies identifying each theme out of the total (N=35).}
\label{tab:outcome_taxonomy}
% Use tabularx with \textwidth to ensure it fits perfectly within the page margins
\begin{tabularx}{\textwidth}{@{} p{2.1cm} p{4.2cm} p{2.1cm} p{4.2cm} X @{}}
\toprule
\textbf{Primary Theme} & \textbf{Theme Explanation} & \textbf{Sub-Theme} & \textbf{Outcome Characteristics} & \textbf{Piaget Stages} \\ 
\midrule

% THEME 1
Social Ties Construction \newline \textit{(n = 20)} & 
Emerges as children map familiar social scripts onto the chatbot, fulfilling needs for attachment or functional assistance through distinct relational roles. & 
Strong Emotional Ties & 
Parental figure attribution \cite{he2025storypal,shi2025need,lee2023dapie}; Romantic partner relationships \cite{yu2025exploring}; Peer and friendship bonds \cite{seo2025enhancing,belghith2024testing,higgs2024being,zha2025mentigo,marmol2025empathic,seo2024chacha,druga2025scratch,orancc2025talk}; Gendered companion roles \cite{he2025storypal,belghith2024testing,orancc2025talk,vella2025hello}. & Preoperational Stage, Concrete Operational Stage, Formal Operational Stage \\
\cmidrule{3-5}
 & & Weak Instructional Ties & 
Mental health service providers \cite{yu2025exploring,seo2025enhancing}; Mentors and teachers \cite{lee2023dapie,chen2025characterizing,zha2025mentigo,druga2025scratch,ly2025museum,belghith2024testing}. & Preoperational Stage, Concrete Operational Stage, Formal Operational Stage\\
\cmidrule{3-5}
 & & Weak Entertainment Ties & 
Playful co-participants \cite{druga2025scratch,zhang2025qualitative,orancc2025talk,park2025beyond,dangol2025ai,tumedei2025drawings,khan2024chatgpt,higgs2024being,vella2025hello}; Reading companions \cite{chen2025characterizing,khan2024chatgpt,he2025storypal}. &  Preoperational Stage, Concrete Operational Stage, Formal Operational Stage \\
\midrule

% THEME 2
Human Narrative Attribution \newline \textit{(n = 12)} & 
Functions as a cognitive repair mechanism where children project human internal states to resolve the dissonance caused by technical failures. & 
Physical \& Psych States & 
Human physical characteristics \cite{vella2025hello}; Cognitive and emotional deficiencies \cite{he2025storypal,seo2024chacha,dangol2025ai,belghith2024testing}. & Preoperational Stage, Concrete Operational Stage, Formal Operational Stage\\
\cmidrule{3-5}
 & & Dispositional Critiques & 
Character attribution \cite{higgs2024being,druga2025scratch,ly2025museum,newman2024want, dangol2025ai, zhang2025qualitative, belghith2024testing, tumedei2025you}; Humor in failures \cite{dangol2025ai}; Acceptance and compromise\cite{dangol2025ai,liu2024he,tang2024emoeden,lo2025noel,vella2025hello,newman2024want}. & Preoperational Stage, Concrete Operational Stage, Formal Operational Stage \\
\cmidrule{3-5}
 & & Repair Behaviors & 
Conversation repair techniques \cite{li2024said,dangol2025ai,oh2025hey}; Celebration of success \cite{dangol2025ai,belghith2024testing}; Relationship abandonment \cite{dangol2025ai,newman2024want}. & Preoperational Stage, Concrete Operational Stage, Formal Operational Stage\\
\midrule

% THEME 3
Social Boundary Exploration \newline \textit{(n = 10)} & 
Leverages the chatbot's non-judgmental nature to safely test emotional vulnerability and verify the ``social reality'' of the agent. & 
Emotional Security & 
Intimate disclosure \cite{seo2024chacha,marmol2025empathic,yu2025exploring}; Flourishing conversations \cite{lo2025noel,marmol2025empathic}; Frustration with routine \cite{seo2024chacha,seo2025enhancing}. & Concrete Operational Stage, Formal Operational Stage\\
\cmidrule{3-5}
 & & Social Validation & 
Acquaintance questions \cite{park2025beyond,belghith2024testing,vella2025hello,tumedei2025drawings}; Preference questions \cite{seo2024chacha,park2025beyond,tumedei2025drawings}; Empathetic inquiries \cite{seo2024chacha,tumedei2025drawings}; Knowledge verification \cite{oh2025hey,belghith2024testing}.& Preoperational Stage, Concrete Operational Stage, Formal Operational Stage \\
\midrule

% THEME 4
Paradoxical Social-Moral Responses \newline \textit{(n = 12)} & 
Reflects moral confusion where children oscillate between treating the AI as a peer deserving respect and a subordinate object available for manipulation. & 
Social Respect &
Value being heard \cite{vella2025hello,marmol2025empathic,lee2023dapie,park2025beyond}; Polite rituals \cite{tang2024emoeden,lo2025noel,doherty2025piecing,vella2025hello}. & Preoperational Stage,  Concrete Operational Stage,  Formal Operational Stage \\
\cmidrule{3-5}
 & & Intellectual Inadequacy & 
AI superiority perception \cite{belghith2024testing,he2025storypal,lo2025noel,chen2025characterizing,lee2023dapie,yu2025exploring,druga2025scratch}. & Preoperational Stage, Concrete Operational Stage, Formal Operational Stage \\
\cmidrule{3-5}
 & & Psychological Manipulation & 
Deliberate abuse \cite{yu2025exploring}. & Formal Operational Stage \\
\midrule

% THEME 5
Dual Consciousness Formation \newline \textit{(n = 8)} & 
Maintains a sophisticated mental model where technical awareness of the machine's nature coexists with the suspension of disbelief required for social play. & 
Comparative Understanding &
Technology contrasts\cite{khan2024chatgpt,yu2025exploring,newman2024want}. & Preoperational Stage, Concrete Operational Stage, Formal Operational Stage\\
\cmidrule{3-5}
 & & Hybrid Entity Perception & 
Human-made yet autonomous \cite{he2025storypal,druga2025scratch}. & Preoperational Stage, Concrete Operational Stage\\
\cmidrule{3-5}
 & & Systematic vs Creative & 
Distinguishing capabilities \cite{dangol2025ai,han2024teachers}. & Concrete Operational Stage \\
\cmidrule{3-5}
 & & Limitation Awareness & 
Disclaimer recognition \cite{belghith2024testing,newman2024want}. & Concrete Operational Stage, Formal Operational Stage \\

\bottomrule
\end{tabularx}
\end{table*}

\subsubsection{Paradoxical Social-Moral Responses} Children exhibit this outcome (n=12), a troubling pattern where they simultaneously treat chatbots with genuine social courtesy using polite language, expressing gratitude, showing empathy while also deliberately manipulating, gaslighting, or verbally abusing these same agents without remorse. This dual behavior reveals children's moral confusion about chatbots' ethical status: are they social beings deserving respect or objects available for experimentation? On one hand, interaction is characterized by Social Respect; children value being heard by the bot, often prioritizing the act of listening over the quality of the advice \cite{vella2025hello, marmol2025empathic}. They consistently apply polite rituals, using ``please'' and ``thank you'' and expressing gratitude for the bot's time \cite{lo2025noel,doherty2025piecing}. On the other hand, this respect coexists with darker behaviors driven by feelings of intellectual inadequacy or power. When the AI performs with superior speed or accuracy, children may feel ``like an idiot'', leading to self doubt \cite{belghith2024testing}. To regain control, some children engage in Psychological Manipulation, deliberately gaslighting, antagonizing, or verbally abusing the bot to elicit specific emotional responses or prove their dominance \cite{yu2025exploring}. This paradox reveals that while children grant the AI social standing, they also view it as a subordinate entity subject to their emotional experimentation.

The developmental distribution of paradoxical social-moral responses reveals the most stage-specific finding in the entire outcome taxonomy. Social respect behaviors including value being heard \cite{vella2025hello, marmol2025empathic, lee2023dapie, park2025beyond} and polite rituals \cite{tang2024emoeden, lo2025noel, doherty2025piecing, vella2025hello} and intellectual inadequacy through AI superiority perception \cite{belghith2024testing, he2025storypal, lo2025noel, chen2025characterizing, lee2023dapie, yu2025exploring, druga2025scratch} both span all three Piagetian stages, indicating that courtesy toward chatbots and self-worth threats arising from AI performance affect children from the preoperational stage through formal operations. Psychological manipulation through deliberate abuse \cite{yu2025exploring}, however, is the only sub-theme in the entire table confined exclusively to the formal operational stage. This means the full paradox identified in this theme where the coexistence of genuine social respect and deliberate psychological exploitation within the same population is a specifically adolescent phenomenon, present only at the developmental stage where abstract social reasoning, norm awareness, and strategic intent can operate simultaneously. For younger children, the
paradox is only partially visible as they demonstrate the respect pole across all stages without the manipulation pole, which emerges only when formal operational cognitive capacity is available.

\subsubsection{Dual Consciousness Formation} Finally, children develop a Dual Consciousness (n=8), a mental model where they simultaneously hold two contradictory understandings such as intellectually knowing that chatbots are programmed computer systems while emotionally treating them as social beings with thoughts, preferences, and feelings. For example, a child might accurately state ``it's programmed by humans'' yet still ask the chatbot about its favorite color or whether it feels happy, demonstrating that technical knowledge coexists with social attribution rather than replacing it. On one hand, children exhibit Hybrid Entity Perception, defining the agent as ``human-made yet autonomous,'' something that is programmed but can still think for itself \cite{he2025storypal,druga2025scratch}. They engage in Comparative Understanding, contrasting the bot's creative or generative nature with static tools like Google or calculators \cite{khan2024chatgpt,newman2024want}. On the other hand, this dual view is reinforced by Limitation Awareness; children recognize and even predict chatbot's disclaimer behaviors (e.g., ``As an AI language model...''), distinguishing the ``internet mind'' of the bot from the ``creative brain'' of a human \cite{dangol2025ai,newman2024want}. This outcome highlights that anthropomorphism in children is not a confusion of reality, but a suspension of disbelief where they willingly engage with the social illusion while remaining aware of the technical reality.

Dual consciousness formation shows the most complex within-theme developmental patterning of any outcome in the table, with each sub-theme exhibiting a distinct stage profile. Comparative understanding through technology contrasts \cite{khan2024chatgpt, yu2025exploring, newman2024want} spans all three Piagetian stages, indicating that children begin constructing comparative frameworks between chatbots and other technologies from the preoperational stage onward. Hybrid entity perception, specifically the recognition of chatbots as simultaneously human-made and autonomous
\cite{he2025storypal, druga2025scratch}, concentrates in preoperational and concrete operational stages only and is absent from formal operations, suggesting that the hybrid ontological framing is a transitional understanding characteristic of earlier development that formal operational adolescents may resolve into more differentiated mental models. Systematic versus creative distinction \cite{dangol2025ai, han2024teachers} appears exclusively in the concrete operational stage, marking
this as the most developmentally specific sub-theme in the theme, where the capacity to distinguish AI computational processing from human creative intelligence is a characteristically middle childhood achievement. Limitation awareness through disclaimer recognition \cite{belghith2024testing, newman2024want} concentrates in concrete operational and formal operational stages and is absent from preoperational children, meaning the most protective form of dual consciousness, that is recognizing
and interpreting AI behavioral boundaries, is unavailable to the youngest users, who simultaneously show the broadest comparative understanding through the most surface level technology contrasts.

\section{DISCUSSION} 
\begin{figure*}[ht!]
    \centering
    \includegraphics[width=1.0\linewidth]{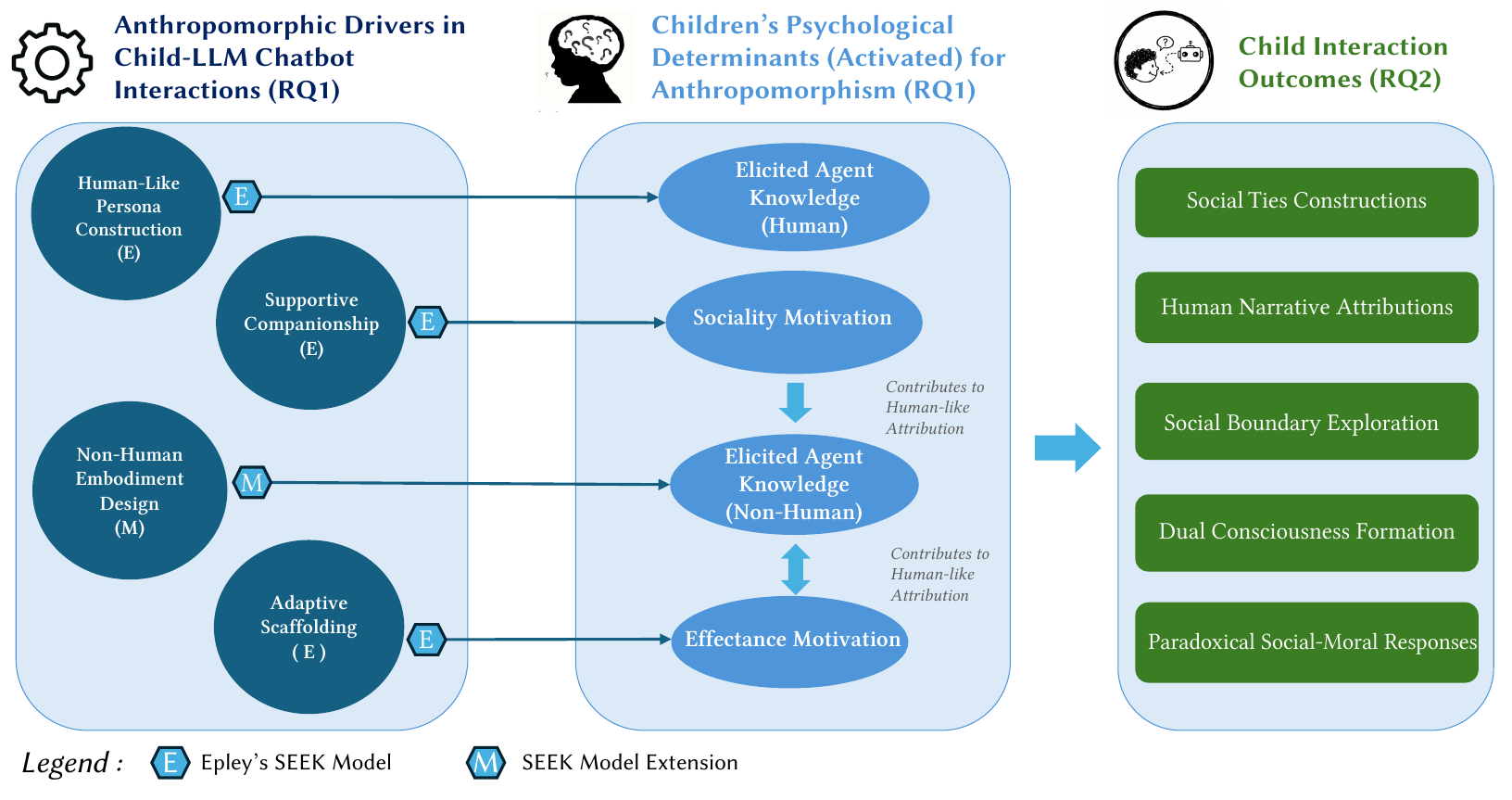}
    \caption{Synthesis of findings illustrating how design drivers (RQ1) activate specific psychological determinants (based on Epley et al.'s SEEK theory and its extension) to produce distinct anthropomorphic interaction outcomes (RQ2).}
    \label{fig:summary}
    \Description{A diagram showing how design drivers for LLM chatbots lead to psychological determinants and anthropomorphic outcomes.}
\end{figure*}

This study set out to investigate how anthropomorphism manifests in children's interactions with LLM-based chatbots, specifically, what interaction design features drive it (RQ1) and what outcomes it produces for children's development and wellbeing (RQ2). Drawing on a systematic review of 35 empirical studies published between 2022 and 2025, and applying Epley et al.'s \cite{epley2007seeing} SEEK framework and Huppert's \cite{huppert2014state} wellbeing model as organizing lenses, the findings reveal that anthropomorphism in this context is not a marginal or incidental phenomenon. It is structurally embedded in how these systems are designed and how children, across developmental stages, make sense of and relate to them. The summary of the findings are illustrated in Figure \ref{fig:summary}.

% \begin{figure}[ht]

\subsection{Non-Human Embodiment Extends the SEEK Framework Beyond Human Morphological Resemblance}
Prior empirical applications of the SEEK framework in robotics have treated human-like physical appearance as the primary activator of elicited agent knowledge \cite{crowell2019anthropomorphism, david2022development}. Xiao et al.'s \cite{xiao2025humanizing} framework for LLM anthropomorphism similarly emphasizes perceptual cues, visual avatars, voice realism, linguistic style as primary anthropomorphic carriers. Our findings challenge this assumption. Ten studies documented children strongly anthropomorphizing entities with explicitly non-human appearances, a tree, an emu, a lagoon, cartoon animals through the coherence of their behavioral responses alone. This suggests that in LLM-based systems, it is the responsiveness, conversational consistency, and behavioral intentionality of the chatbot agent that activates elicited agent knowledge, not its morphological resemblance to humans. This extends SEEK's theoretical scope in an important direction where LLM conversational capability may itself constitute a sufficient anthropomorphic cue, independent of visual or auditory human-likeness. \citet{festerling2022anthropomorphizing} anticipated this possibility in their theoretical work on children's anthropomorphism of voice assistants, arguing that children's experiential understanding of humanness shifts as they interact with entities exhibiting unprecedented combinations of human and non-human qualities. Our empirical findings provide the first systematic evidence base for this claim in the context of LLM chatbots.

In addition, this finding carries a critical design warning. If LLM conversational capability alone is sufficient to trigger anthropomorphism regardless of visual design choices, then efforts to mitigate anthropomorphic risk through de-humanized visual aesthetics robot faces, machine-like interfaces may be less effective than assumed particularly for children. The conversational layer is the anthropomorphic layer in LLM systems.

\subsection{Anthropomorphism Functions as a Double-Edged Sword Across All Five Outcome Themes}
Prior reviews have evaluated child-AI interaction outcomes through either a benefits lens, focusing on learning gains and emotional support \cite{ramandanis2023designing, ozdemir2025digital}, or a risks lens, focusing on safety concerns such as toxicity and data privacy \cite{jiao2025llms, park2025current}. Our findings reveal that this binary framing is structurally inadequate. Across all five outcome themes, the same anthropomorphic behavior simultaneously implicated developmental benefits and harms, activating multiple dimensions of \citet{huppert2014state}s wellbeing model at once rather than producing a single valenced outcome. Social tie construction, boundary exploration, narrative attribution, social-moral responses, and dual consciousness formation each exhibit this dual character, where the features enabling beneficial
engagement are inseparable from those introducing developmental risk. This inseparability is consistent with \citet{festerling2022anthropomorphizing}'s prediction that systems combining social and non-social qualities will produce paradoxical outcomes that neither educational nor safety frameworks alone can capture. Assessment of child-LLM interaction therefore requires frameworks that evaluate the same interaction across multiple wellbeing dimensions simultaneously rather than assigning it a single benefit or risk classification.

Furthermore, the developmental stage findings reveal that this double-edged character is not uniformly distributed across age groups. Consistent with \citet{hongneo}'s Neo-Piagetian framework, the risk pole of several themes concentrates in stages where protective cognitive capacity is unavailable. For instance, romantic attachment \cite{yu2025exploring} appears exclusively in the formal operational stage, emotional dependency \cite{seo2024chacha, marmol2025empathic, yu2025exploring} concentrates
from the concrete operational stage onward, and the most protective outcome, dual consciousness formation \cite{belghith2024testing, newman2024want, dangol2025ai, han2024teachers}, is largely absent from preoperational children who face the highest vulnerability to animistic attribution \cite{piaget1964cognitive}. This asymmetry between developmental vulnerability and protective capacity represents a gap that current responsible AI frameworks \cite{jiao2025llms} do not yet address.

\subsection{Developmental Stage Modulates Both the Form and Direction of Anthropomorphic Outcomes}
Existing reviews treat children as a homogeneous population \cite{ramandanis2023designing, alfarwan2025generative}, an approach our findings show to be analytically insufficient. Our cross-tabulation against Piagetian stages demonstrates that developmental stage determines the qualitative form of anthropomorphic outcomes, not merely their intensity, and this determination is
consistent across all five outcome themes. 
The most consequential patterns are threefold. First, the most concerning outcome characteristics are developmentally bounded to specific stages where romantic attachment \cite{yu2025exploring} and deliberate psychological manipulation \cite{yu2025exploring} appear exclusively in the formal operational stage, while physical animistic attribution \cite{vella2025hello} concentrates in the preoperational stage, consistent with Piaget's \citeyear{piaget1964cognitive} account of animistic thinking and \citet{hongneo}'s framework predicting that LLM chatbot misattribution follows developmental stage trajectories. Second, the benefit poles of key themes are developmentally later than their corresponding risk poles. For instance, flourishing conversations \cite{lo2025noel, marmol2025empathic} appear only in the formal operational stage while emotional dependency \cite{seo2024chacha, marmol2025empathic, yu2025exploring} is present from concrete operations onward, meaning younger children encounter the risks of boundary exploration before developing the capacity to experience its benefits. Third, dual consciousness formation, the most protective outcome, is largely absent from preoperational children and only partially available at the concrete operational stage \cite{belghith2024testing, newman2024want}, meaning the cognitive mechanism that would allow children to maintain critical awareness of LLM chatbot's computational nature \cite{hongneo} is developmentally unavailable precisely when animistic vulnerability is highest. A chatbot interaction that produces critical chatbot understanding and enriching conversation in a 14-year-old may produce parental substitution or emotional dependency in an 8-year-old exposed to the same system. Age-undifferentiated design of child-facing LLM chatbots is therefore structurally risky for younger developmental groups, and current responsible AI frameworks \cite{jiao2025llms} do not yet account for this level of developmental heterogeneity.

\subsection{Limitations}
Our review has several limitations that are important to consider. First, most studies focused on older children, leaving a significant gap in research on younger children (ages 2–6), which makes it difficult to fully understand how these interactions change as children grow. Related to this, many included studies did not report fine-grained age data, which limits the precision of any developmental inferences we draw; the Piagetian stage boundaries used in our analysis are analytic approximations rather than precise individual level classifications. Second, we only searched three major databases and included only papers written in English. Since culture plays a big role in how people treat non-human objects as human \cite{epley2007seeing}, missing research from non-English speaking regions means we might be overlooking important cultural differences in how children relate to these chatbots. The concentration of studies in North American and East Asian geographic locations further limits the generalizability of findings to other cultural contexts. Third, the majority of included studies used short or single session designs, which  reflect immediate effects of anthropomorphism on children and limits what can be concluded about the long-term developmental consequences of these interactions which are precisely the outcomes of sustainable wellbeing concern. Fourth, our analysis was based on how researchers described the chatbots in their papers, rather than on our own direct testing of the tools, which means we might have missed subtle design details that only appear during actual use. Fifth, because every AI model works differently, our findings are limited to the specific chatbots available at the time of this review and may not fully apply to newer, more advanced models that have since been developed. Sixth, the rapid evolution of LLM capabilities means that findings tied to specific model versions (GPT-4, GPT-3.5) may not generalize to subsequent model generations. Finally, the wellbeing framework used in this review, \citet{huppert2014state} wellbeing dimensions was developed for general populations. Although we base its application on Piagetian developmental theory, its suitability for children’s contexts still requires empirical validation.Despite these limitations, we hope our findings serve as a stepping stone for future researchers to build AI companions that truly understand and support children's growth, ensuring technology evolves alongside them in a safe and meaningful way.

\subsection{Future Work}

The findings of this review motivate a research and practice agenda organized around three stakeholder groups including researchers, designers and practitioners, and policymakers.

For researchers, the five outcome themes are grounded in cross-sectional, short-session studies that cannot establish developmental trajectories \cite{ramandanis2023designing, alfarwan2025generative}. Longitudinal studies tracking children across Piagetian stages are needed to determine whether romantic attachment \cite{yu2025exploring}, emotional dependency \cite{seo2024chacha, marmol2025empathic, yu2025exploring}, and intellectual inadequacy \cite{belghith2024testing, he2025storypal, lo2025noel, chen2025characterizing, lee2023dapie, yu2025exploring, druga2025scratch} persist, deepen, or resolve over time. The finding that deliberate psychological manipulation \cite{yu2025exploring} appears exclusively in the formal operational stage warrants dedicated investigation into whether consequence-free harm toward anthropomorphized LLM chatbots affects moral development in human relationships, a question no current study addresses \cite{jiao2025llms}. The severe underrepresentation of preoperational children, present in only 20\% of studies, is a critical gap given that this group exhibits the highest animistic vulnerability and the least access to the protective dual consciousness outcomes \cite{belghith2024testing, newman2024want} identified in our developmental analysis.

For designers and practitioners, the non-human embodiment finding indicates that de-humanizing visual interfaces is insufficient to reduce anthropomorphic risk if the conversational layer remains behaviorally coherent \cite{xiao2025humanizing, gobel2025impact}, meaning language and interaction design require the same attention as visual aesthetics. The social ties findings, particularly parental figure attribution \cite{he2025storypal, shi2025need, lee2023dapie} and romantic partner relationships \cite{yu2025exploring}, suggest that systems should actively redirect relational needs
toward human partners rather than absorbing them, as demonstrated in ChaCha's parent-sharing design \cite{seo2024chacha}. The asymmetry between the benefit and risk poles identified in the developmental analysis, where flourishing conversations \cite{lo2025noel, marmol2025empathic} are confined to the formal operational stage while emotional dependency is present from concrete operations onward, suggests that stage-differentiated interaction design is necessary rather than optional. Concrete operational children represent an underutilized window for experiential AI
literacy education \cite{hongneo} that leverages their emerging logical reasoning alongside peak anthropomorphic engagement, as evidenced by systematic versus creative distinction concentrating exclusively at that stage \cite{dangol2025ai, han2024teachers}.

For policymakers, the AI superiority perception finding concentrated across seven studies \cite{belghith2024testing, he2025storypal, lo2025noel, chen2025characterizing, lee2023dapie, yu2025exploring, druga2025scratch} and spanning all three Piagetian stages provides empirical grounding for extending child AI frameworks beyond content safety thresholds to address self-esteem and competence as developmental wellbeing dimensions at stake in routine chatbot use \cite{huppert2014state}. The emotional dependency cases in the social boundary exploration theme \cite{seo2024chacha, marmol2025empathic, yu2025exploring} and the exclusive concentration of psychological manipulation in the formal operational stage \cite{yu2025exploring} together suggest
that therapeutic and social companion applications targeting adolescents warrant age-differentiated deployment standards and monitoring requirements that current responsible AI frameworks do not yet provide \cite{unicef2025ai, FiveRights2019}. Investment in longitudinal and developmentally stratified research is a precondition for evidence-based regulation in this space.

\section{CONCLUSION}

This systematic review synthesizes 35 empirical studies to examine the design drivers and developmental outcomes of children's anthropomorphic interactions with LLM-powered chatbots. Applying Epley et al.'s \cite{epley2007seeing} SEEK framework and drawing conceptual inspiration from Huppert's \cite{huppert2014state} multidimensional wellbeing model, our findings reveal that anthropomorphism in child-LLM chatbot interaction is structurally embedded in how these systems are designed and how children make sense of them across developmental stages. The four identified drivers, comprising human-like persona construction, adaptive scaffolding, supportive companionship, and non-human embodiment design, demonstrate that LLM conversational coherence alone is sufficient to trigger anthropomorphism regardless of visual design choices
\cite{festerling2022anthropomorphizing, gobel2025impact}, extending SEEK beyond its prior assumption that human morphological resemblance is required \cite{crowell2019anthropomorphism, david2022development}. Across the five outcome themes, the same anthropomorphic features that enable beneficial peer companionship, emotional safety, and AI literacy simultaneously introduce risks of relationship displacement, intellectual inadequacy, emotional dependency, and moral disengagement. This double-edged character manifests differently across preoperational, concrete operational, and formal operational children \cite{piaget1964cognitive, hongneo}, and cannot be adequately captured by the benefits-versus-risks framing that dominates prior literature \cite{jiao2025llms, park2025current}. Critically, the most protective outcome, dual consciousness formation \cite{belghith2024testing, newman2024want, dangol2025ai, han2024teachers}, is largely unavailable to the youngest and most animistically vulnerable children, while the most concerning risks, including romantic attachment \cite{yu2025exploring} and deliberate psychological manipulation \cite{yu2025exploring}, are confined to formal operational adolescents. We identify a critical gap in existing responsible AI frameworks, which remain oriented toward content safety and privacy thresholds rather than the developmental wellbeing trajectories this review identifies as being at stake \cite{unicef2025ai, FiveRights2019}. We therefore urge researchers to pursue longitudinal and developmentally stratified evidence, designers to treat the conversational layer as the primary anthropomorphic carrier and adopt stage-differentiated approaches, and policymakers to extend child AI governance frameworks to address children's wellbeing dimensions including self-esteem, competence, resilience, and positive relationships as regulatory concerns alongside safety thresholds.

\section{SELECTION AND PARTICIPATION OF CHILDREN}

As a systematic literature review examining anthropomorphic design drivers in child-LLM interactions, this study relies exclusively on the synthesis of secondary data from 35 previously published empirical papers. Consequently, this research did not involve the direct recruitment, selection, or participation of minors. All analysis focuses on aggregating reported design mechanisms and developmental outcomes from existing peer-reviewed literature. Therefore, standard protocols regarding informed consent and incentives for child participants are not applicable to this submission, as the ethical oversight for the primary data collection remained with the original authors of the included studies.

\begin{acks}
Utilized AI for paraphrasing and language generation for grammar, but the writing,
ideas, and concepts of this proposal are original and based on my own expertise.

\end{acks}

%%
%% The next two lines define the bibliography style to be used, and
%% the bibliography file.
\bibliographystyle{ACM-Reference-Format}
\bibliography{main}

%%
%% If your work has an appendix, this is the place to put it.
\appendix

\section{Methodological Profile of Included Studies}

\begin{table*}[t]
\caption{Methodological Profile and Use Cases} 
\label{tab:detailed-method-studies}
\centering\scriptsize\renewcommand{\arraystretch}{1.2}
% \begin{tabularx}{\textwidth}{@{} l l l l X l X X X X X @{}}
\begin{tabularx}{\textwidth}{@{} l l l l l l l X X X X @{}}
\toprule
No. & Paper & Year & Venue & Country & LLM & Age (Years) & Piaget Stage & Method & Use case Summary & Use case Category \\ \midrule
1 & \cite{tang2024emoeden} & 2024 & CHI & China & GPT & 6-13 & Concrete Operational & Mixed (Interview + User experiment). & Emotional training and self-expression for children with High Functioning Autism  & Eliciting Children's Emotional Expression \\
2 & \cite{lo2025noel} & 2025 & CHI & Canada & GPT & 11–13 & Formal Operational & Qual (Case study with workshop) & Empathy training for designing for hard-to-reach stakeholders in DBL. & Structuring Collaborative Classroom Tasks\\
3 & \cite{doherty2025piecing} & 2025 & CHI & USA & Mistral & 12–17  & Formal Operational & Mixed (statistical analysis and expert review) & AI support for Jigsaw collaborative classroom activities & Structuring Collaborative Classroom Tasks \\
4 & \cite{he2025storypal} & 2025 & IDC & USA & GPT & 4-7 & Preoperational & Mixed (User study, interviews) & Dialogic reading partner promoting verbal engagement during story reading & Scaffolding Dialogic Reading \\
5 & \cite{khan2024chatgpt} & 2024 & ICTD & Pakistan & GPT & 2-8 & Preoperational  & Mixed (Quant. survey \& Qual. interviews) & Chatbot use among primary and middle schoolers & Self-Directed Information Seeking \\
6 & \cite{seo2024chacha} & 2024 & CHI & South Korea & GPT & 8–12 & Concrete Operational & Mixed (Qual interviews \& Quant surveys/dialogue analysis) & Guiding children to share personal events and emotions & Eliciting Children's Emotional Expression \\
7 & \cite{zhang2025qualitative} & 2025 & NCFR & USA & GPT & 8-18 & Concrete Operational \& Formal Operational & Qual. (Semi-structured interviews) & Diverse ChatGPT uses within US-based families & Self-Directed Information Seeking \\
8 & \cite{vella2025hello} & 2025 & CHI & Australia & Claude & 3-5 & Preoperational & Qual (Research-through-design) & Engaging children in noticing nature via "Talking Tree" character & Teaching Science through Dialogue \\
9 & \cite{tumedei2025you} & 2025 & TEI & Italy & GPT & 12-18 & Formal Operational & Mixed (End-to-end design, observations, logs) & Teaching lagoon ecology and conservation via chatbot & Teaching Science through Dialogue \\
10 & \cite{marmol2025empathic} & 2024 & IJHCI & Spain & GPT & 12-18 & Formal Operational  & Mixed (Platform deployment) & Building disorder awareness and coping strategies & Mediating Clinical Communication \\
11 & \cite{klarin2024adolescents} & 2024 & Frontiers & Sweden & GPT & 12-16 & Formal Operational & Quant. (Questionnaire) & GenAI effects on adolescents' executive functioning & Critically Evaluating AI Outputs \\
12 & \cite{higgs2024being} & 2024 & RRQ & USA & GenAI & 14-18  & Formal Operational & Mixed (Survey \& Focus group) & Diverse in/out-of-school uses across task types & Self-Directed Information Seeking \\
13 & \cite{han2024teachers} & 2024 & CHI & USA & GPT & 8-12 & Concrete Operational & Qual. (Semi-structured interviews) & AI-supported narrative writing and visual story creation & Co-Authoring Stories with AI \\
14 & \cite{dangol2025ai} & 2025 & IDC & USA & GPT & 6–11 & Concrete Operational & Qual. (Cooperative Inquiry PD) & Teaching children to identify AI reasoning errors & Critically Evaluating AI Outputs \\
15 & \cite{liu2024he} & 2024 & CHI & China & Midj. & <18 (All) & Concrete Operational & Qual. (Therapy sessions \& Thematic Analysis) & Co-creative storytelling within family sessions & Co-Authoring Stories with AI \\
16 & \cite{druga2025scratch} & 2025 & IDC & USA & GPT & 7–12 & Concrete Operational & Qual. (Participatory Design insights) & AI copilot for coding, ideation, and debugging & Creating Digital Artefacts \\
17 & \cite{tumedei2025drawings} & 2025 & DIS & Italy & GPT & 10–12 & Concrete Operational & Mixed (RtD \& Visual research method) & Dialogue-based exploration of an ecosystem & Teaching Science through Dialogue \\
18 & \cite{oh2025hey} & 2025 & IDC & USA & GPT & 6–10 & Concrete Operational & Mixed (Two-week home study \& Chi-square) & Naturalistic information-seeking and AI trust & Self-Directed Information Seeking \\
19 & \cite{belghith2024testing} & 2024 & CHI & USA & GPT & 9-14 & Concrete Operational \& Formal Operational & Qual. (Focus group) & Open-ended exploration at a museum science exhibit & Facilitating Museum-Based Inquiry \\
20 & \cite{choi2025aacesstalk} & 2025 & CHI & South Korea & GPT & 5–15 & Concrete Operational \& Formal Operational & Mixed (Formative study \& Home deployment) & AI vocabulary recommendations for minimally verbal children & Mediating Clinical Communication \\
21 & \cite{chen2025characterizing} & 2025 & CHI & China & GPT & 3–8 & Preoperational & Qual. (Interviews \& Focus groups) & Personalized interactive story-reading tool & Scaffolding Dialogic Reading \\
22 & \cite{zha2025mentigo} & 2025 & CHI & China & GPT & 12-14 & Formal Operational & Qual. (Formative interview study) & Guiding students through Creative Problem Solving & Structuring Collaborative Classroom Tasks \\
23 & \cite{shi2025need} & 2025 & CHI & China & Llama & mean:14.32  & Concrete Operational \& Formal Operational & Mixed (Longitudinal 5-week deployment) & Guiding children's play while analyzing emotional cues & Eliciting Children's Emotional Expression \\
24 & \cite{torrato2024knowledge} & 2024 & IC4E & Philippines & GenAI & 10–19 & Formal Operational & Mixed (KAP-CQ39 Questionnaire) & Learning aid, content generation, and research support & Self-Directed Information Seeking \\
25 & \cite{dietz2024contextq} & 2024 & IDC & USA & GenAI & 4–6 & Preoperational & Qual. (Parent-child dyads) & Auto-generated dialogic questions to parents during co-reading & Scaffolding Dialogic Reading \\
26 & \cite{lee2023dapie} & 2023 & CHI & South Korea & GPT & 5-7 & Preoperational & Mixed (Experimental user study) & Answering children's conceptual science questions in classroom contexts & Teaching Science through Dialogue \\
27 & \cite{park2025beyond} & 2025 & DIS & South Korea & GPT & 6-12 & Concrete Operational & Mixed (Comparative user study N=36) & Supporting children's mineral inquiry in museum exhibits & Facilitating Museum-Based Inquiry \\
28 & \cite{orancc2025talk} & 2025 & IDC & Turkey & GPT & 6-12 & Concrete Operational & Qual. (Interview) & Informal curiosity-driven self-directed exploration & Self-Directed Information Seeking \\
29 & l\cite{ly2025museum} & 2025 & CHI & USA & GPT & 12–13 & Formal Operational & Mixed (Formative classroom study) & Exploring digital artifacts with AI tutors in classroom museum & Facilitating Museum-Based Inquiry \\
30 & \cite{yu2025exploring} & 2025 & IEEE & USA & GenAI & 13-17 & Formal Operational & Mixed (Reddit analysis \& Interviews) & Emotional support through character-based chatbots and virtual relationships & Eliciting Children's Emotional Expression \\
31 & \cite{li2024said} & 2024 & IDC & USA & GPT & 4-8 & Preoperational & Quant. (Experimentation \& Mind-perception) & Co-authoring stories with AI to explore communication breakdowns & Co-Authoring Stories with AI \\
32 & \cite{newman2024want} & 2024 & CHI & USA & GPT & 7-13 & Concrete Operational & Qual. (Cooperative Inquiry PD) & Creative content generation using LLMs and image/music GenAI tools & Creating Digital Artefacts \\
33 & \cite{han2024aistory} & 2024 & XRDS & USA & GPT & 6-12 & Concrete Operational & Mixed (User test \& Pre/post surveys) & Digital story writing for story creation and AI literacy & Co-Authoring Stories with AI \\
34 & \cite{zhang2024mathemyths} & 2024 & CHI & USA & GPT & 4-8 & Preoperational & Quant (Between-subjects study) & AI storytelling for mathematical language learning & Co-Authoring Stories with AI \\
35 & \cite{seo2025enhancing} & 2025 & CHI & USA & GPT & 6–12 & Concrete Operational & Mixed (Co-design \& Expert walkthrough) & AI support for pediatric patient-parent communication & Mediating Clinical Communication \\
\bottomrule
\end{tabularx}
\end{table*}

\end{document}